\documentclass[sigconf]{acmart}
\usepackage{xspace}
\usepackage[nolist,nohyperlinks]{acronym}
\usepackage{xcolor}
\usepackage{url}
\usepackage{xurl}
\usepackage{subcaption}
\usepackage{enumitem}

\copyrightyear{2020}
\acmYear{2020}
\setcopyright{licensedusgovmixed}\acmConference[IMC '20]{ACM Internet Measurement Conference}{October 27--29, 2020}{Virtual Event, USA}
\acmBooktitle{ACM Internet Measurement Conference (IMC '20), October 27--29, 2020, Virtual Event, USA}
\acmPrice{15.00}
\acmDOI{10.1145/3419394.3423615}
\acmISBN{978-1-4503-8138-3/20/10}

\newif\ifcomment
\commenttrue
\ifcomment

\newcommand{\jbnote}[1]{{\bf \textcolor{magenta}{[JB: #1]}}}
\newcommand{\ernote}[1]{{\bf \textcolor{teal}{[ER: #1]}}}
\newcommand{\rbnote}[1]{{\bf \textcolor{orange}{[RB: #1]}}}

\else
\newcommand{\jbnote}[1]{}
\newcommand{\ernote}[1]{}
\newcommand{\rbnote}[1]{}

\fi

\graphicspath{{figs/}}

\newcommand{\etal}{\emph{et al.}\xspace}
\newcommand{\ie}{i.e.\xspace}
\newcommand{\eg}{e.g.\xspace}
\newcommand{\dis}{Dissenter\xspace}
\newcommand{\yt}{YouTube\xspace}
\newcommand{\punkt}[1]{\item\textbf{\emph{#1}}}

\setlist[itemize]{leftmargin=*}

\begin{document}

\title{Reading In-Between the Lines: An Analysis of Dissenter}

\author{Erik Rye}
  \affiliation{
    \institution{CMAND}
  }
  \email{rye@cmand.org}

\author{Jeremy Blackburn}
  \affiliation{
    \institution{Binghamton University}
  }
  \email{jblackbu@binghamton.edu}

\author{Robert Beverly}
  \affiliation{
    \institution{Naval Postgraduate School}
  }
  \email{rbeverly@nps.edu}

\begin{abstract}
Efforts by content creators and social networks to enforce legal and
policy-based norms, \eg blocking hate speech and users,
has driven the rise of unrestricted communication platforms.  One such
recent effort is \dis, a browser and web application that provides a conversational
overlay for any web page.  These conversations hide in plain sight --
users of \dis can see and participate in this conversation, whereas
visitors using other browsers are oblivious to their existence.  Further, the website and
content owners have no power over the
conversation as it resides in an overlay outside their control.

In this work, we obtain a history of \dis comments, users,
and the websites being discussed, from the initial release of \dis in
Feb.\ 2019 through Apr.\ 2020 (14 months). Our corpus consists of approximately
  1.68M comments made by 101k users commenting on 588k distinct URLs. We first analyze macro
characteristics of the network, including the user-base, comment
distribution, and growth.  We then use toxicity dictionaries, Perspective API, and a Natural Language Processing
model to understand the nature of the comments and measure the
propensity of particular websites and content to elicit hateful and
offensive \dis comments.  Using curated rankings of media
bias, we examine the conditional probability of hateful comments given
left and right-leaning content. Finally, we study \dis as a social
network, and identify a core group of users with high comment toxicity.
\end{abstract}



\maketitle

\section{Introduction}
\label{sec:intro}

Virtual communities and discussion permeate modern society, and have
fundamentally changed the way information is disseminated and
consumed.  Platforms that support such information exchange face not
only technical challenges, but also legal and policy-based content
concerns.  Recently, major platforms have established and enforced
policies to restrict hate speech \cite{twitter-policy} and users that engage in
it. Participants of these communities have therefore moved to
new platforms that either passively tolerate or actively support this
content.  For example \cite{zannettou2018gab,8508809} examine 
speech and political characteristics of Gab, 
while \cite{waseem2016hateful} characterizes hate speech in Twitter.

\dis began as a web browser plugin, but, after
being banned from the major browsers' respective plugin and extension
stores~\cite{dissenter-removed}, morphed into a self-contained, full-fledged browser based
on Brave~\cite{dissenter-github,brave}.
Characterizing itself as the ``free speech web browser,'' \dis provides, 
for any URL, a tightly
integrated discussion forum specific to that URL.  Only \dis users see
this discussion forum.  Thus, while \eg a news site may have its own
discussion forum for a particular article, \dis provides a parallel
universe where its community of users are free to discuss (presumably
a dissenting opinion) without restriction.  Notably, the website and
content owners have no power over this discussion forum as it
resides in an overlay outside their control.

Similar forms of web annotation and augmentation have been created in the past, \eg Google
Sidewiki~\cite{sidewiki} (now defunct) and Hypothesis~\cite{hypothesis}.
These efforts, however, were launched in an era pre-dating
restrictions on social
media content and not aimed at freedom of speech or 
providing a platform for fringe groups to discuss particular websites
and content.
In our work, the first to attempt to measure and 
characterize \dis, we obtain a history of \dis comments,
users, and the websites being discussed from the initial release of 
\dis in Feb.\ 2019 through Apr.\ 2020 (14 months). 
We find more than 101k \dis users contributing more than 1.68M comments
on 588k unique URLs.

Given recent debate surrounding censorship and the 
role of social media platforms in society -- with the United States
President signing executive orders to prevent censorship -- our work is especially
timely~\cite{covfefe}.  Toward a deeper understanding of \dis as an
emergent platform, 
we make the following contributions:
\begin{itemize}
 \item Characterization of the \dis user base, including the 
       intersection with Gab and Reddit
 \item Analysis of the \dis social network, including
       influential users and the set of users within individual
       comment threads.
 \item Classification of the toxicity of \dis comments and correlation
   of classes with both the political bias of the \acp{URL} being
       commented on, as well as its content.
\end{itemize}

\section{\dis}
\label{sec:background}

In this section, we describe \dis, its relationship to its
parent application, Gab, and discuss major changes \dis
has undergone since its launch in early 2019. We then define
\dis-specific terminology.

\subsection{History}

A description of the \dis plugin, browser, and comment overlay system,
necessarily begins with Gab~\cite{gab}. Gab was founded by
Andrew Torba in 2016 as an alternative social network to more mainstream
platforms; Gab counts among its users figures banned from Twitter such as
provocateur Milo Yiannopoulos, and reality television star Tila Tequila. While
Gab's stated purpose is to ``champion free speech, individual liberty and the
free flow of information online''~\cite{gab}, studies~\cite{zannettou2018gab,
lima2018inside} suggest Gab is primarily a fringe social network that contains 
hate speech and extremist content. Gab received significant attention
after anti-Semitic hate speech was
discovered on the Gab account of the individual responsible for the
2018 Tree of Life synagogue shooting in Pittsburgh~\cite{treeoflifecase,zannettou2020merchantsICWSM}.

\dis, billed as the ``Comment Section of the Internet'', was released by
Gab in February 2019 as a reaction to the disabling of ``comment sections'' of
some websites by content providers, including YouTube. A full 25\% of \dis users
we examine in this study refer to ``censorship'' in their profile's biography,
suggesting that perceived censorship is a common motivator for much of \dis's
user base.
In order to restore the
ability for Internet users to comment on web content, \dis acts as a
comment aggregation platform, receiving comments pertaining to \acp{URL} and
displaying them to other users of its service. In this manner, \dis acts as a
kind of overlay, displaying this ``hidden'' content only to users of
\dis,
while visitors that do not use the application remain unaware of
its existence.

\begin{figure}[t]
 \centering
 \resizebox{1.0\columnwidth}{!}{\includegraphics{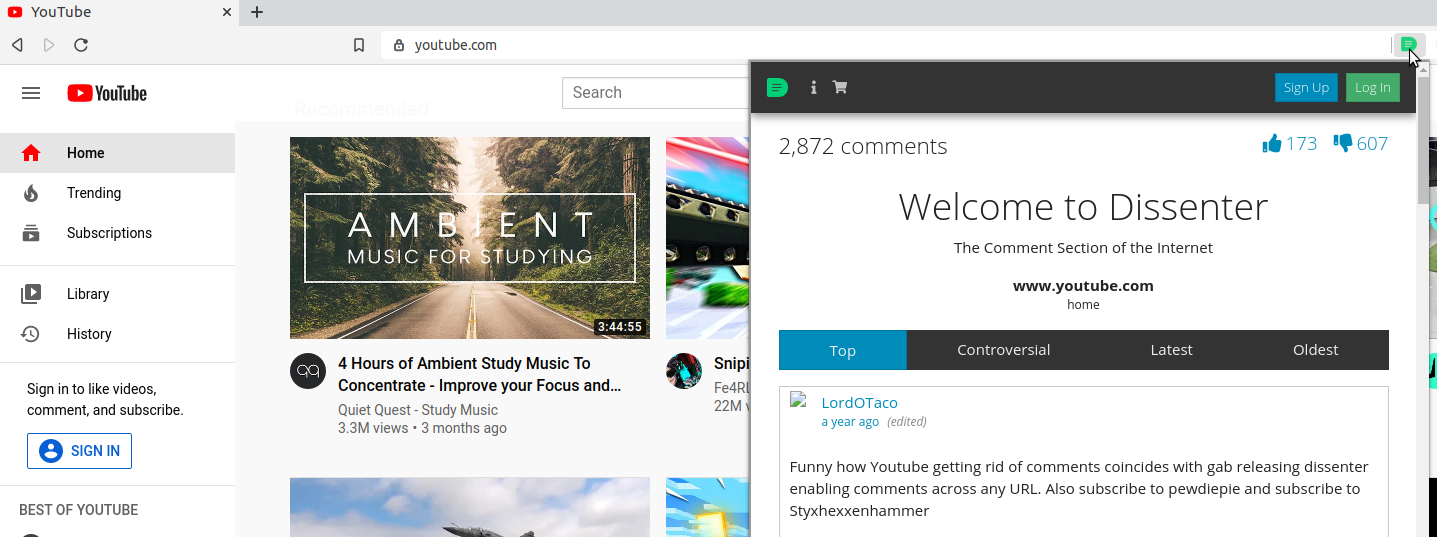}}
 \vspace{-5mm}
 \caption{\dis browser view of YouTube}
 \label{fig:dissenter}
 \vspace{-5mm}
\end{figure}

\dis initially took the form of a Firefox and Chrome browser
extension which, when toggled, allowed users
to post
and view comments for a given \ac{URL}.
In April 2019, only two months after launching, the \dis
extension was removed by both the Mozilla and Chrome extension stores.
Both stores cited a terms of service
violation, claiming that the extension 
was used to post hate speech~\cite{dissenter-removed}.
\dis then morphed into a standalone browser by forking the
Brave web browser~\cite{brave,dissenter-github}.
Figure~\ref{fig:dissenter} shows an
example of the conversation overlay visible when viewing the YouTube
home page using the \dis browser.

While providing a standalone browser extricated the \dis
plugin from oversight by corporations that might otherwise attempt to crack down
on the speech of its users, it requires users to switch
their default web browser.  

To augment the \dis browser, and provide a second method to access
\dis comments, Gab deployed a news aggregation site
called Gab Trends in October 2019~\cite{gab-trends}.  
Gab Trends
presents titles and short
summaries of news articles from around the web, and includes 
\dis comment threads for each article.  The comment thread visible
via the \dis browser and Gab Trends is identical.  Registered \dis users can 
participate in the \dis discussion by using the Trends web portal.
Further, the Trends home page allows submission of new \acp{URL}.
Upon submitting the new
\ac{URL}, the user is directed to a web page containing all of the \dis comments
that have previously been  made about this \ac{URL}; if the \ac{URL} is new to
the \dis and Gab Trends system, this page contains no comments, but allows new
users that navigate to it to make comments about this \ac{URL}. 

The popularity of Gab Trends is disputed.  Gab itself claims
3M monthly views~\cite{torba_views}, while independent sources
estimate 67k unique views per month~\cite{examiner_views}.

\subsection{Terminology}
\label{sec:dissenter_terms}

Analogous to other social networks, \dis users have a
\emph{home page}.
A home page
lists their username (a unique handle \eg ``\texttt{@a}''),
display name (which may
differ from the username, \eg ``Andrew Torba''),
a biographical statement, and
a profile image. Importantly, home pages list all of the \acp{URL} the
user has commented upon. 

Each \ac{URL} that has received a \dis comment or been entered into \dis
has a \emph{comment page}. A comment page is
analogous to a home page for a particular \ac{URL}. Each comment page contains a
title, which generally corresponds to the title content in the HTML of the page the
\ac{URL} points to, and a brief description of the content, which is typically
generated by the first paragraph at the underlying \ac{URL}. Some exceptions to
title and description content exist, particularly when the content being
commented upon is from another social network. For instance, both YouTube and Twitter
content is typically embedded in the comment page by the \dis system, leading to ambiguous or
altogether absent titles and descriptions. In order to make up for this lack of
content data from \dis itself, we handle YouTube separately, which we describe in
\S\ref{sec:addl_scrape}. Each comment page contains all of the \dis comments that
have been made pertaining to the \ac{URL}, and all of the replies to those
comments. 

While uncovering the operation of \dis, we find several 
undocumented identifiers within their HTML and JavaScript.
We use these unique 12 byte \dis identifiers 
to prevent duplication and ensure uniqueness of users
and content.
Each user has a unique 24 hexadecimal digit
\emph{author-id}. Similarly, each distinct \ac{URL} in \dis 
has a 12 byte 
\emph{commenturl-id} identifier.
Finally, every comment and reply is also assigned a 12 byte 
\emph{comment-id}.

We discovered that these identifiers are not entirely random or a
hash, but rather contain some structure.  Analyzing the identifier,
we find that all three encode state about their
creation time.
The first 4 bytes of the
\emph{author-}, \emph{commenturl-}, and \emph{comment-ids} are a Unix timestamp
in seconds that describes the creation of entity; for example, an account
created on February 28, 2019 at 16:23:53 UTC, will have an \emph{author-id} beginning with
\texttt{5c780b19}.
Similarly, a \emph{commenturl-id} encodes the first time a URL appears in \dis.
While there appears to be
additional structure in the remaining 16 hexadecimal digits, we are unable to
determine its meaning as of this writing. In order to verify these findings, we
created our own \dis accounts and posted innocuous content.

Finally, when a user posts a comment or reply in \dis, they have
the option to label it as \ac{NSFW}. By default, these posts
are invisible both to unauthenticated and authenticated \dis users; in order to
view this content, a logged-in user must explicitly ``opt-in'' via the \dis
settings page. Because \ac{NSFW} posts are hidden from all but
authenticated users that have opted-in,
this effectively creates hidden content
within a shadow overlay. Similarly, an ``offensive'' label also exists for \dis
comments, although
unlike \ac{NSFW}, it is not tagged by the user creating the content. As with
\ac{NSFW} content, an authenticated user must opt-in to viewing ``offensive''
comments.

\section{Methodology}
\label{sec:method}

The primary interface to \dis is their browser or website.  While \dis
has an API, it is neither documented nor intended for public use.
Hence, our methodology required basic reverse engineering of their
platform and the application of several techniques to completely and
programmatically gather \dis-internal data including users, user
meta-data (including the social network), comments, and the \acp{URL}
for each comment thread.  Using this methodology, we effectively
mirror the \dis database.

This section first details our data collection campaign and shows the
steps taken to verify its accuracy and completeness.  We then augment
the \dis data by gathering the content of selected \acp{URL} commented
upon for additional context.  Finally, we describe our comment content
classification techniques to enrich our understanding of comment
content and user behavior.  Ethical considerations of our work 
are provided in \S\ref{appendix:ethics}. 

\subsection{Gab-Based Username Harvesting}
\label{sec:gab-users}

While \dis home pages provide important data on users and links to
comments, there is no publicly available central database of
usernames.  Thus, a central component of our methodology is to harvest
\dis usernames.  
Note that registering for \dis requires an active Gab account.  Therefore, 
\dis users are necessarily Gab users, and we leverage this fact in order to
enumerate \dis users before beginning to crawl other \dis content.

Initially, we attempted to gather Gab usernames via a combination of mining
Pushshift.io~\cite{baumgartner2020pushshift} and crawling the most popular Gab
account's (``\texttt{@a}'', belonging to Gab founder Andrew Torba) followers, which is
automatically followed by new users on the platform when their account is
created. However, this methodology failed to uncover users that had not posted on
Gab or had manually ceased following \texttt{@a}, and our results suggested a
period of time before the \texttt{@a} handle was automatically followed by new
users.

\begin{figure}[t]
 \centering
 \resizebox{0.8\columnwidth}{!}{\includegraphics{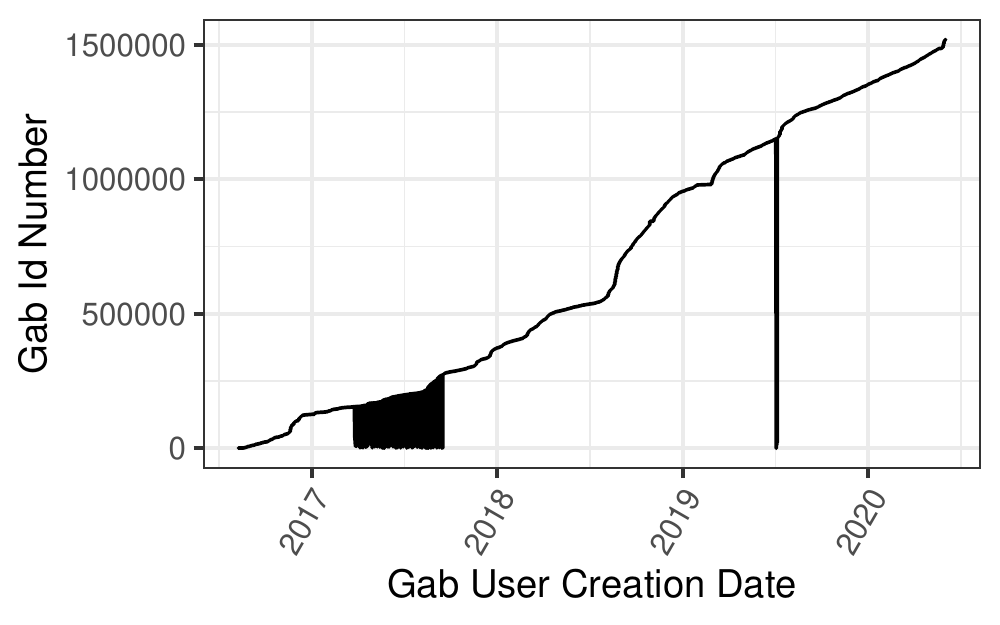}}
\vspace{-3mm}
 \caption{Gab User IDs Assigned to New Accounts Over Time}
 \label{fig:gabIds}
\end{figure}

As discussed in \S\ref{sec:dissenter_terms}, each \dis user account is
associated with a unique identifier. In a similar vein, Gab accounts also have a
unique user identifier. Unlike \dis's \emph{author-id}s, however, Gab user IDs
do not encode creation time, but are instead a counter beginning at 1, the user
ID associated with ``\texttt{@e}'', belonging to former Gab \ac{CTO} Ekrem
B\"uy\"ukkaya. Having created a test account for which the Gab ID is known, we query the
Gab API endpoint \url{https://gab.com/api/v1/accounts/<Gab ID>} for IDs
between 1 and our account's ID to retrieve JSON-encoded information pertaining
to that user. Gab's API helpfully returns an error when an ID is not associated
with a user account, and in this manner, we are able to exhaustively enumerate
Gab's user base. Among the information present in the user data JSON is the
account creation date and time, which largely confirms the hypothesis that the
Gab ID is a monotone increasing counter. Some exceptions to monotonicity exist
in which Gab assigned unallocated, lower-valued ID numbers to new
user accounts; whether these ID values became free after a user deleted their
account, or whether gaps were deliberately placed between consecutively
allocated IDs earlier in its history is unclear. Figure~\ref{fig:gabIds} shows
when the account associated with each Gab ID number was created; apart from two
distinct time periods in mid-2017 and mid-2019, Gab IDs are generally assigned
sequentially and are strictly increasing.

This enumeration process reveals 1.3M distinct accounts, a significant number
more than discovered in prior work in 2018~\cite{zannettou2018gab}, which
discovered 336k users. In addition to the two years separating
\cite{zannettou2018gab} and our study, the difference is explained by a large
number of discovered Gab users that have not posted any messages, do not follow
any other Gab users, and are similarly not followed by any other users. While
many or most of these accounts may belong to inactive users, we find several
thousand ``friendless'' and ``silent'' Gab users that are otherwise active on
\dis. Finally, because our Gab user discovery methodology enumerates all Gab
users, our \dis results in \S\ref{sec:results} do not suffer from questions of
completeness or sensitivity that a user harvesting strategy based on spidering a
high-degree user's followers might.

Next, we determine which Gab users are also \dis users. For each Gab username
discovered, we send an HTTP request to the \ac{URL} of the corresponding \dis
home page, if it exists (\url{https://dissenter.com/user/<Gab username>}).
Based on the HTTP response sizes, we are able to identify \dis accounts, which
are at least 10 kB; responses for non-existent users are $\sim$150 bytes.
Of the 1.3M Gab usernames we enumerated via its API, 101k also have \dis
accounts, representing approximately $8\%$ of all Gab users.

\subsection{Dissenter Comment Harvesting}
\label{sec:dissenter_scrape_method}

With the usernames discovered in  
\S\ref{sec:gab-users}, we crawled
\dis for \acp{URL} that users comment on, the comments
they made about those pages, as well as replies to other users'
comments.  Our crawler 
first visits the home page of each \dis user to 
capture their meta-data, including
username, display name, \emph{author-id}, and biography.  Then
the crawler gathers the set of \acp{URL} the user has commented on.

We then iterate over the set of
commented-upon \acp{URL}. For each \ac{URL}, we visit its comment page
in \dis and 
collect the \emph{commenturl-id}, 
the number of comments and number of up-
and down-votes the \ac{URL} has received, as well as the title and
brief description. As noted in \S\ref{sec:dissenter_terms}, the title and
description may be ambiguous or empty, depending on the underlying content the
commented-upon \ac{URL} describes and the ability of the \dis system to parse
this data. Within each comment page, we iterate over the comments and
replies.
For each comment, we record the \emph{author-id},
\emph{comment-id}, 
and the comment text. Comment text appears to have no character limit;
the longest comment we find is $>$90k characters, consisting of the word
``ha'' repeated 45k times in response to a YouTube video discussing Facebook's
political bias. Replies can be made in response to both comments as well as to
replies themselves; there also appears to be no practical limit on the
depth at which replies might be made, \eg, a reply to a reply to a reply is
valid. In addition to the data we collect for
comments, we also note the \emph{comment-id} of the content to which the reply
is replying.
Although the HTTP response headers indicate that a rate-limit of 10 requests per
minute is employed by the servers hosting \dis content, this counter is
\emph{per-\ac{URL}}; because we do not need to request the same \ac{URL} twice
in our crawl, we are unimpeded by this rate-limit.

Each comment is available from
the \ac{URL}
\url{https://dissenter.com/comment/<CID>/}, 
where
\texttt{<CID>} is the \emph{comment-id}. We call this page the 
\emph{comment-page}, as its purpose is to display a single comment 
as well as any replies.
For reasons that are unclear, comment pages contain a JavaScript element 
with an unused (commented-out) JavaScript variable
called ``\texttt{commentAuthor}''. \texttt{commentAuthor} defines an array 
with user data.  While much of this embedded user data is identical
to what is available to us via the user's home page, it 
also contains otherwise undiscoverable meta-data including the user's
language setting, permissions, and view-filter preferences.  We save these additional
hidden meta-data as part of our per-user characterization.

To obtain the \ac{NSFW} and ``offensive'' content described in
\S\ref{sec:dissenter_terms}, we re-spider \dis using the HTTP cookies of an
authenticated account we created with \ac{NSFW} and ``offensive'' content enabled
separately, so that we are able to discern between content labeled \ac{NSFW} by
the submitter and comments marked as ``offensive''.
These comments have no specific flag or other identifier present in the document
body to indicate their presence;
therefore, we infer \ac{NSFW} and ``offensive'' comments as those found when
authenticated with these flags enabled that were not previously discovered. In
order to ensure we do not erroneously mislabel content as \ac{NSFW} or
``offensive''
because of crawler errors, we monitor request timeouts and re-request missed
pages, and ensure that subsequent runs only consider comments made during the
initial spider's time frame. Additionally, we manually confirm 100
comments classified as \ac{NSFW} or ``offensive'' by our subsequent crawls by
attempting to view these comments both while authenticated with the \ac{NSFW}
and ``offensive'' view preferences enabled and while not authenticated. Our
\ac{NSFW} and ``offensive'' comment results are discussed in
\S\ref{sec:nsfwcomments}.

In total, we obtain 1.68M comments and replies on 588k distinct \acp{URL} made
by $>$101k users via our methodology.  Macro characteristics of these
data are discussed in detail in \S\ref{sec:results}.

\subsection{YouTube Crawling}
\label{sec:addl_scrape}

Typically, we rely on the title and description provided by the \dis
application in the comment page in order to gain valuable context about
\acp{URL} being commented upon. However, \dis's own methodology for \ac{URL}
content appears unable to handle the most popular source of commented-upon
\acp{URL}: YouTube videos. These pages generally appear
with the title "/watch" and a null description, although the video
itself is embedded in the page. Therefore, because YouTube
content in particular represents a sizable percentage of our data (128k
\acp{URL}) and because we seek to understand the content that generates 
comments,
we also gather the content of the underlying web page for YouTube \dis comments.
Because YouTube pages require JavaScript to render properly,
we use Selenium~\cite{selenium} to automate content retrieval. 
The data we seek (\eg, video title, uploader name)
resides in large blocks of JavaScript, which may explain its absence from
\dis.  For each \ac{URL}, we classify the content as one of three distinct types --
``video'', pages that contain a single \yt video, ``user'', a home page for a
particular \yt user, and ``channel'', which is a collection of videos under a
single banner. 

\subsection{Social Network Crawling}
\label{sec:friends}

Finally, we return to Gab in order to gain context about the social network that
\dis users inhabit. 
Social relationships on Gab are directional; much like in Twitter, a user
may become a follower of another user, and may accumulate followers themselves.
While \dis users are able to ``follow'' other \dis users,
none of the \dis browser, plugin, user home page, web application, or hidden
meta-data reveal followers, or allow even an authenticated
user to view his or her followers and
following users.  Presumably this is because the social network aspect
of \dis is a subset of Gab, and is an as-yet unimplemented part of the
\dis experience.
Therefore, we use Gab followers as a proxy, and because Gab users are a strict
superset of \dis users, any two \dis users can follow each other on Gab. 

We use the Gab API in order to obtain these relationships for further analysis
in \S\ref{sec:results}. 
Using the Gab API, we gather the followers and followed users of each \dis user.  
We note that Gab exposes its rate-limiting in the HTTP response
headers by including 
the number of remaining
requests, as well as the time at which the request limit will be
refreshed. 
To minimize impact on the service,
we issue at most one request per second, and monitor the number of
remaining requests. If necessary, we wait until the number of available requests
has been refreshed before continuing to issue new requests for Gab friends.
Note that results from querying the Gab API for the social network are 
paginated, thus we can ensure that we gather the complete network
graph.

Finally, by removing non-\dis users from the followers and those
followed obtained by querying Gab,
we construct a \dis-specific social network graph.  

\subsection{Classification}
\label{sec:classification}

To gain a more complete understanding of \dis, we must understand the
content and context of the comments and replies.  In particular, we
are interested in assessing the degree of toxicity and offensiveness.
While significant prior work exists on automatically labeling hate and
toxic speech, current approaches yield accuracies between
70-80\%~\cite{ICWSM1715665} and it remains an open research problem.  
For example, there are indications that the
models encode racial bias~\cite{sap2019risk}, while some models
can be deceived~\cite{hosseini2017deceiving}.  

To
underscore the difficulty of the problem, consider an innocuous
comment about the country Pakistan.  This comment could be construed
as hateful as it contains the substring ``paki,'' a false positive.
However, not performing stemming and fuzzy matching can yield false
negatives, for instance if the hate word is succeeded with a ``z''
when using slang.  Words themselves are ambiguous and must be taken in
full context.  For example, the term ``skank'' can be used as a hate
term or in reference to a style of dance. 

Because our work is focused on characterizing \dis rather than
improving the state-of-the-art in hate speech detection, we therefore
explored multiple approaches to label comments in order to bound our
estimates of its toxicity.

\begin{itemize}
\punkt{Dictionary} We utilize the modified Hatebase~\cite{hatebase} dictionary of toxic
terms used by the authors of~\cite{hine2017kek}
and~\cite{zannettou2018gab}.  This dictionary contains 1,027 hate
words.  We tokenize each \dis comment and reply, perform stemming, and
then count the number of tokens that match a term in the dictionary.
Our per-comment hate dictionary score is then the ratio of hate words
over the number of tokens in the comment.  While this metric is
simple, it misses important context in the comments.  For instance,
the ambiguous terms ``queen'' and ``pig'' appear in the dictionary.
However, by using the same dictionary as these prior works, we can
draw direct meaningful comparisons.

\punkt{Perspective} Next, we leverage the Google Perspective API.
The Perspective API provides several models that provide scores for
different aspects of toxicity.  Perspective allows us to effectively
outsource comment scoring; however, as with the other methods, it
has limitations.  
The API is
trained primarily on Wikipedia data, and thus there are some questions
about its portability.

\punkt{NLP} Finally, we employ Natural Language Processing models to build a
three-class (hate, offensive, or neither) comment classifier.  To
train our classifier, we use labeled data from~\cite{ICWSM1715665}
which contains 1,194 hate, 16,025 offensive, and 20,499 neither labels
of Twitter tweets gathered via crowd-sourcing.  Because of the
imbalanced complexion of data, we use ADASYN to
oversample~\cite{he2008adasyn}.
 
We experiment with neural networks, decision trees, and support vector
machines (SVMs) using 1 and 2-grams of cleaned and stemmed
word tokens.  Using grid search to tune the hyperparameters, we
achieve the highest accuracy using SVMs.  With 5-fold
cross-validation, we achieve an F1 score of 0.87 on the Twitter
training dataset.  Using this SVM model, we compute the probability of
each of the three possible classes for all \dis comments and replies.
\end{itemize}

\begin{figure}[t]
 \centering
 \resizebox{0.8\columnwidth}{!}{\includegraphics{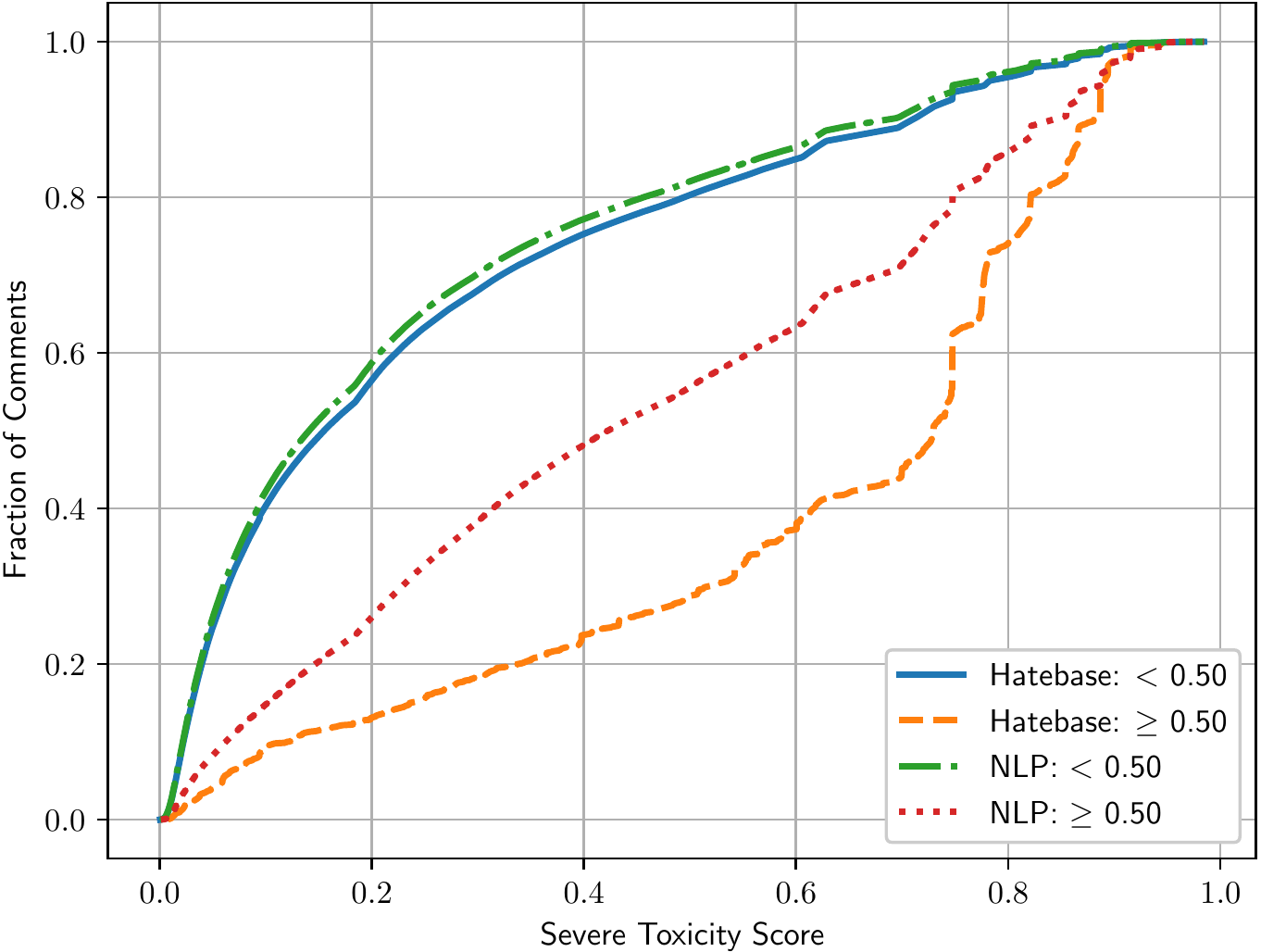}}
\vspace{-3mm}
 \caption{Comparing Dictionary and NLP Classifiers Against Perspective}
 \label{fig:toxic-v-hate}
\vspace{-3mm}
\end{figure}

To better understand the differences of each classifier as applied to 
\dis data, we evaluate all comments and replies with the three approaches
and compare the resulting scores in Figure~\ref{fig:toxic-v-hate}.  We
see that the classifiers largely agree -- the distribution of
Perspective toxicity for comments that score low with Hatebase or the
NLP classifier is significantly skewed toward less toxicity, while 
those with high scores are skewed toward higher Perspective toxicity.

For our purposes, issues surrounding toxicity classification are
somewhat mitigated.  First, we are less interested in scoring any
\emph{particular} comment, and instead are interested in aggregate
trends and the distribution of scores.  Second, we compare \dis to
several baselines which gives us an idea of \emph{relative}
differences in scores across users and communities.

Recent efforts by Google have sought to make Perspective's performance
more transparent~\cite{perspective2019}, while third-parties have
evaluated perspective on canonical datasets and found it to exhibit
high precision and recall~\cite{pavlopoulos2019convai}.  As such,
Perspective provides a publicly available platform that has been
independently validated.  Because of the general agreement between
classifiers and our use of the classification as a relative metric, we
use Perspective for the remainder of our evaluation.

\section{Results}
\label{sec:results}

This section begins with a high-level characterization of the \dis
platform, analyzing users and \acp{URL} being discussed.  
We then measure hate speech and toxicity in
\dis as compared to other platforms, as well as investigate the
relationship of toxicity to content and the social network of
\dis users.

\subsection{\dis Users}
\label{sec:user_results}

\subsubsection{How popular is \dis?}

We first examine the set of \dis users we discover using the methodology
of~\S\ref{sec:gab-users}.
As noted in \S\ref{sec:dissenter_terms}, the
\emph{author-id} identifier encodes
each account's creation time. 
\dis experienced
a steep initial influx of users to the platform, as nearly 79k (77\%) joined through the
first full month of operation (March, 2019).

\begin{figure}[t]
 \centering
 \resizebox{0.8\columnwidth}{!}{\includegraphics{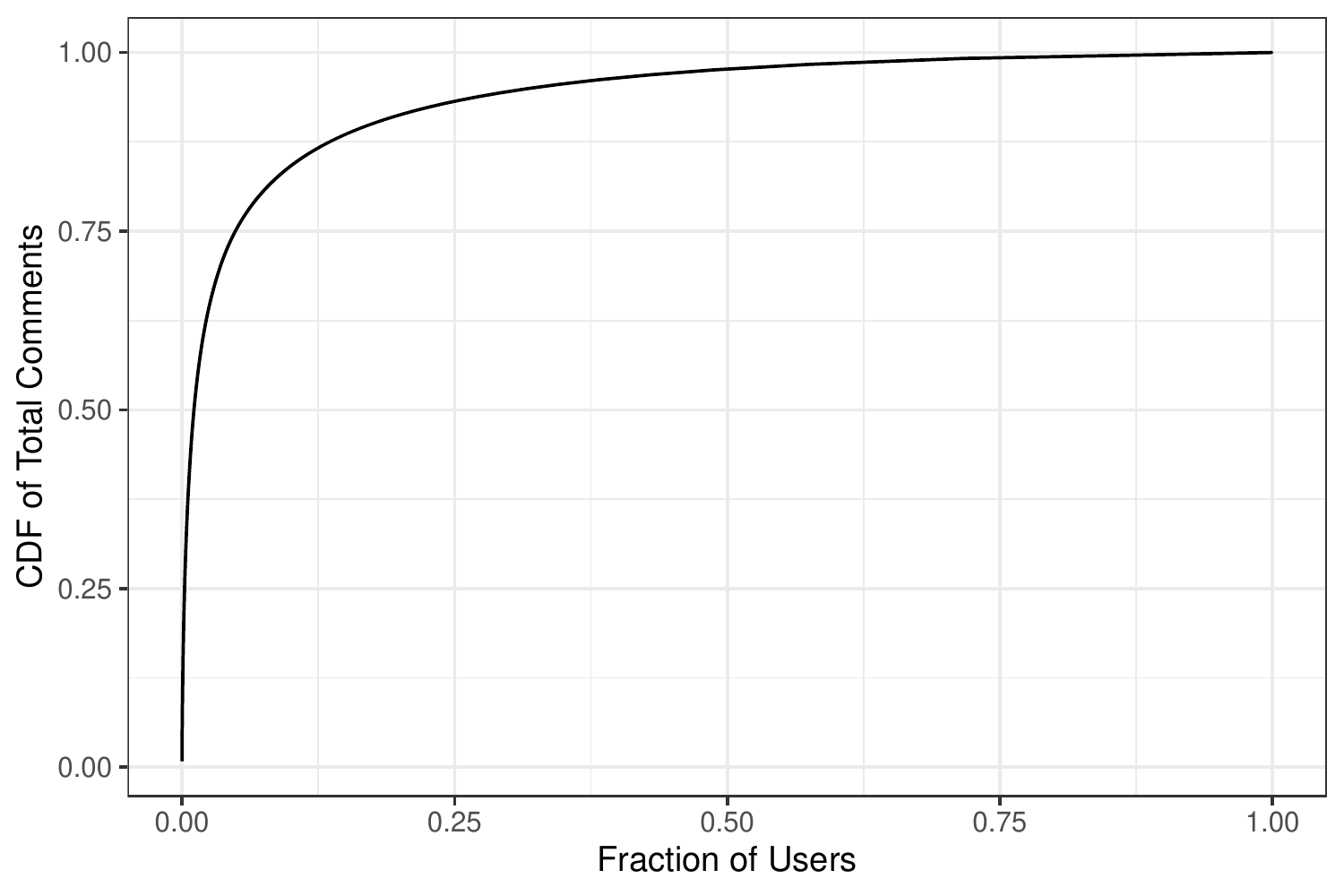}}
 \vspace{-3mm}
  \caption{\dis Comments and Replies per Active User}
 \label{fig:commentcount_cdf}
 \vspace{-3mm}
\end{figure}

Of the more than 101k unique usernames we discover, approximately 47k (47\%) 
commented on at least one \ac{URL}. Considering only these active users that
have made at least one comment, Figure~\ref{fig:commentcount_cdf} shows that
approximately 90\% of comments are made by about 14\% of active users (7\% of
total users). The long tail of Figure~\ref{fig:commentcount_cdf} indicates that
many users made a relatively small number of comments. 
We note that none of the top users
fall into the top twenty Gab users by number of followers, score, or PageRank as
determined by prior work~\cite{zannettou2018gab}, nor are they
prominent in the \dis social network as will be shown
in~\S\ref{sec:results:toxicity}. Finally, we discover approximately 1,300 users
who commented on \acp{URL} through our \dis crawl that did not appear in our
enumeration of Gab's users in \S\ref{sec:gab-users}. This finding was
surprising,
as an active Gab account is a prerequisite for creating a \dis
account. On closer examination, we discovered that these accounts appeared to be
deleted by their owners, as their Gab home pages matched the appearance of an
account that we test-deleted. Interestingly, these users' \dis accounts and
comments remain despite the deletion of their Gab account.

\textbf{Takeaways: } 
Slightly more than half (53\%) of \dis's users have not commented on a \ac{URL}
or replied to another user's comment. This does not necessarily mean that these
users are inactive; users can interact with \dis by giving ``thumbs up'' or
``thumbs down'' on both the \ac{URL} and other users' comments, but these
actions are transparent to us.

While \dis's user base is a strict subset of Gab's, \dis is not
simply a Gab in miniature.
Its core group of users are extremely active on the
site, posting thousands of comments in little over a year on web content that they
presumably consume beforehand. 

We discover more than 1,300 users whose Gab accounts were deleted. The comments
left by these users remain on \dis, and because their Gab account no longer
exists, they are unable to authenticate %
to delete these posts.

\subsubsection{User Characterization}
Using the %
embedded JavaScript data described in
\S\ref{sec:dissenter_scrape_method}, we are able to more extensively
characterize the 47k active users.
Two
\dis users are flagged as ``isAdmin'': \texttt{@a}, Andrew Torba's account,  and
\texttt{@shadowknight412}, which belongs to Rob Colbert, the Gab CTO.
Despite the existence of ``isModerator'', no active
accounts we queried had this moderator flag set, although it is possible one or
more moderator accounts exist that do not post comments. Eight accounts were banned from the
platform; of these, several had no obvious explanation for
being banned based on the comments we
obtained, while for others the reason is clear. For example, one account is clearly related to
a home remodeling company and posted only advertisements for its business, while
another posted what appears to be the home address of a federal official and
expressed a desire for an ``accident'' to happen to that individual.
Table~\ref{table:usersettings} counts the frequency of all possible 
account flags across the active user set.

Users can further apply filters to show or hide comments on \acp{URL}
based on categories of users or comment labels. For instance, ``\ac{NSFW}'' is a
label that a user may apply when posting a comment, and ``pro'' is a label
that a designates a paid GabPRO account -- a status that unlocks
additional platform features, such as the removal of ads and ability
to upload larger videos.
Table~\ref{table:usersettings} summarizes the number of positive responses for
each comment filter. Of note, nearly all users choose to see content from
``pro'', ``verified'', and ``standard'' \dis accounts; this is unsurprising, as
all three of these are applied by default. However, the ``\ac{NSFW}''
and ``offensive'' preferences are by default disabled. We discover 18k comments
classified by \dis as \ac{NSFW} ($\sim$10k) or ``offensive'' ($\sim$8k),
though we were unable to determine with certainty what moderation policies or
user feedback generates an ``offensive'' classification. Because only 15\%
and 7\% of active users enable the \ac{NSFW} and ``offensive'' filters,
respectively, these comments constitute a kind of shadow platform \emph{within}
\dis.

\textbf{Takeaways: } Artifacts in the \dis HTML source show that even in \dis,
users get banned for speech that is deemed unacceptable. Further, the ``view''
flags indicate that \ac{NSFW} and offensive content exists as a shadow overlay on the \dis
overlay itself, viewable only to a small fraction of \dis users.
\begin{table*}[t]
  \caption{User Attribute Flags and Comment View-Filters Enabled for Active
  Users ($n$=47,165)}
  \label{table:usersettings}
  \vspace{-3mm}
\small{
\begin{tabular}{|l|r|l|r|l|r||l|r|}
\hline
\multicolumn{6}{|c||}{\textbf{User Flags}} & \multicolumn{2}{c|}{\textbf{Comment Filters}} \\ \hline
  canLogin & 47,152 (99.97\%) & isBanned  & 8 (0.02\%) & is\_investor & 137
  (0.29\%) & pro & 47,093 (99.85\%)\\ \hline
  canPost & 47,150 (99.97\%) & isAdmin & 2 (0.00\%) & is\_premium & 61 (0.13\%)
  & verified & 47,103 (99.87\%) \\ \hline
  canReport & 47,158 (99.99\%) & isModerator & 0 (0.00\%)& is\_tippable & 73
  (0.15\%) & standard & 47,112 (99.89\%) \\ \hline
  canChat & 47,153 (99.97\%) & is\_pro & 1,257 (2.67\%)& is\_private & 1,838
  (3.90\%) & nsfw & 7,094 (15.04\%)\\ \hline
  canVote & 47,152 (99.97\%) & is\_donor & 397 (0.84\%) & verified & 485
  (1.03\%) & offensive & 3,456 (7.33\%)\\ \hline
\end{tabular}
}
  \vspace{-2mm}
\end{table*}

\subsection{Content Analysis}
\label{sec:url_results}

Since \dis serves users concerned that their ability to comment directly on the source material might be curtailed, it is only intuitive to examine what content they discuss.

\subsubsection{What URLs are being commented on?}

We discover 588k  \acp{URL} that have been commented upon according to
\dis's own unique identifier, the \emph{commenturl-id}.  However, this number
over-counts unique content in two ways. First, \dis differentiates between
\acp{URL} that differ only in the protocol portion; that is, the HTTP and HTTPS
version of \acp{URL} will receive different \emph{commenturl-ids}, separate
comment pages, and can contain entirely different comment content. We observe
400 distinct \acp{URL} that differ only in the protocol part of the \ac{URL};
another 60 differ only by the presence or absence of a trailing forward-slash
character. %
Second, \dis's handling of \acp{URL} with HTTP GET query parameters
causes unique content over-counting. Many of the \acp{URL} we
observe contain several GET parameters separated by the ``\texttt{\&}'' character; however,
because page content is typically only determined by a single parameter, if at
all, it is likely unnecessary to store more than the first key-value pair as
part of the \ac{URL} in the \dis system.

\begin{table*}[t]
 \centering
  \caption{Most Frequently Commented \acp{TLD} and Domains}
  \vspace{-3mm}
\label{table:tlds_domains}
\small{
\begin{tabular}{|l|r|l|r||l|r|l|r|}
\hline
  \multicolumn{4}{|c||}{\textbf{Top Level Domains}}
 & \multicolumn{4}{c|}{\textbf{Domain}} \\ \hline
.com & 455,885 (77.57\%) & .au & 6,892 (1.17\%) & youtube.com & 121,928 (20.75\%) & foxnews.com & 12,196 (2.08\%) \\ \hline
.uk & 43,808 (7.45\%)& .ca & 5,490 (0.93\%) & twitter.com & 40,392 (6.87\%) & bitchute.com & 12,124 (2.06\%) \\ \hline
.org & 19,502 (3.32\%) &.net & 4,787 (0.81\%) & breitbart.com & 23,705 (4.03\%) & zerohedge.com & 8,634 (1.47\%) \\ \hline
.de & 10,257 (1.75\%) &.nz & 2,979 (0.51\%) &  bbc.co.uk & 16,213 (2.76\%)& theguardian.com & 8,010 (1.36\%)  \\ \hline
.be & 8,013 (1.36\%) &.no & 2,928 (0.50\%) &  dailymail.co.uk & 15,752 (2.68\%) & youtu.be & 7,819 (1.33\%) \\ \hline

Other & 27,194 (0.05\%) & \textbf{Total} & 587,735 (100\%) & Other & 320,962
  (54.61\%) & \textbf{Total} & 587,735 (100\%)\\ \hline
\end{tabular}
}
  \vspace{-2mm}
\end{table*}

A full 97\% (571k) of URLs in Dissenter are HTTPS; another
2\% (15k) are HTTP, and a small fraction contain browser-specific protocols, such
as \texttt{chrome://}. Thirteen \acp{URL} contain the
\texttt{file} protocol, indicating that the \ac{URL} points to a file on the
user's file system. While most of these \acp{URL} point to Windows ``letter
drives'' like \texttt{C:\textbackslash}, several include file paths that appear
to point to legitimate documents on the user's file system.

Of the \acp{URL} we discover within \dis, the overwhelming majority
point to pages under the \texttt{.com} \ac{TLD} (78\%); the
second-most-frequent \ac{TLD} is \texttt{.uk} (7.5\%).
While \texttt{.be} rounds out the top five \acp{TLD},
it most frequently appears a domain hack for \yt \acp{URL} (\eg
\texttt{youtu.be/id}) rather than for Belgian content. 

In addition to the popular \acp{TLD}, 
Table~\ref{table:tlds_domains} gives the most popular second-level
domains by percentage of \acp{URL}.
\yt
is by far the most common,
comprising about 21\% of all \acp{URL} we discovered with comments between the
\texttt{youtube.com} and \texttt{youtu.be} domains. Twitter content is the
second most frequent at about 7\% of all \acp{URL}.
With the exception of Bitchute, a video hosting alternative to \yt oriented 
towards the same general user base as Gab~\cite{trujillo2020bitchute}, the remainder 
of the top \acp{URL} are news-oriented websites. In contrast, when we rank
domains by median comment volume per \ac{URL}, \yt ranks very low, with a median
comment count of 1. Domains with the highest comment volumes per URL are
typically fringe content with a small number of commented \acp{URL}. For
instance, \texttt{thewatcherfiles.com}, a conspiracy aggregation site, ranks
first with 116 comments on one \ac{URL} about the Jewish Blood Libel; the second
highest comment volume domain is \texttt{deutschland.de}, with 95 comments on a
single \ac{URL}, most of which express anger about the Muslim diaspora in Europe.

\textbf{Takeaways: } \dis comments are typically made on video streaming, social
media, or news sites, with \yt comprising the largest fraction of commented
domains. However, domains with the highest comment volume per \ac{URL} are
disjoint from the set of highest commented \ac{URL} counts and often contain
fringe content. \dis users can comment on \emph{any} \ac{URL}, including
those that are local to their own file system or are non-existent.

\subsubsection{\yt}

\yt content comprises a large fraction of the \acp{URL} commented on in
\dis; Table~\ref{table:tlds_domains} shows that $\sim$22\% of \acp{URL}
we obtained with comments are \yt content. Further, \dis
comment-pages typically contain little information about the video itself,
likely because this information is dynamically generated and thus difficult for
\dis itself to mine. This creates difficulty in understanding the content at
the \ac{URL}, and is compounded by the fact that \yt videos also have no
inherent bias, making broad generalizations of this content
effectively impossible. 

Therefore, as described in~\S\ref{sec:addl_scrape}, we gather and
analyze the
content of 128k \yt \ac{URL} present in our data.
The majority of these \yt \acp{URL} are videos: 125k
are labeled video content, along with 2k channels and 1k users. Because \yt
videos may be taken down by the owner or the platform itself for a variety of
reasons, we discovered only 109k active video pages, while 16k were unavailable.
While the most common reason for video removal was a generic ``Video
Unavailable'' label, 3k videos were listed as private and required permission to view,
another 3k were unavailable because the \yt account that had posted them was
terminated, and nearly 400 were removed for violating \yt's hate speech policy. It is noteworthy that
even in the event that \yt takes action to remove objectionable comment, \dis
still provides a platform for users to comment on what \emph{was} at that
\ac{URL}, serving as an ersatz digital history for the content that once existed
there.

Each active video has a ``content-owner'', the name of the entity or individual
that uploaded the content.  
Interestingly, Fox News and CNN -- generally considered to be on opposite
ends of the ideological spectrum -- both appear in the top six most
commented upon \yt content producers. 2.4\% of all \yt videos that
had a comment were produced by Fox News, as compared with 0.6\% for CNN.
Normalized by the fraction of all videos produced by each news source,
4.7\% of all Fox News videos have at least one \dis comment while only 
0.5\% of CNN videos were commented upon.
Slightly more than 10\% of the active videos we
crawl have their comment functionality disabled on the \yt platform, reinforcing
\dis's argument that it provides an outlet for users to express their opinions
on content where it would otherwise not be allowed. 

\textbf{Takeaways: } \yt is a sizable fraction of all \dis
comment \acp{URL}; the videos \dis users comment on frequently disallow
commenting, and are often removed from \yt altogether for a variety of reasons.

\subsubsection{Are there language differences in \dis comments?}
Using the \texttt{langid}~\cite{langid-github} language identification tool, we
classify each of the 1.68M comments and replies in our dataset. Our results
indicate that \dis comments are overwhelmingly in English (1.57/1.68M
or 94\%).
German is
the second most popular language with 31k (2\%); this matches our
expectations as \texttt{.de} is the fourth most-common \ac{TLD} and first
non-English-speaking country's \ac{ccTLD} in Table~\ref{table:tlds_domains}. French,
Spanish, and Italian complete the top five most-frequent languages, each with
less than 0.5\% popularity.

\textbf{Takeaways: } The vast majority (94\%) of \dis comments are in English, with
German the only other language achieving $>$1\% representation.

\subsection{\dis Toxicity}

The concept of toxicity online has gained a bit of a spotlight lately.
In a nutshell, toxicity is loosely defined as anti-social behavior that causes harm to a community at the social level.
Things like harassment, hate speech, personal attacks, and trolling can all be
considered toxic, as they reduce the inherent utility of the platform they occur
on, as well as harm its underlying community of users.
Previous work~\cite{zannettou2018gab,8508809}, as well as articles and world events have indicated that Gab is more toxic than the average community.

In fact, at least part of the motivation behind Gab and Dissenter's existence is that its user base was considered too toxic for platforms like Twitter, and that the discussions they have are similarly considered unsuitable by many platforms.
This raises several interesting questions that we aim to answer in this section
using our content classification techniques
in~\S\ref{sec:classification}.
Are \dis users particularly toxic?
What kind of toxicity is exhibited?

\subsubsection{NSFW and Offensive Comments}
\label{sec:nsfwcomments}

As described in \S\ref{sec:dissenter_terms}, a user is presented with the
option to label a comment or reply as \ac{NSFW} when posting. This label
prevents the comment from appearing to any user that has not explicitly opted-in
to seeing this content. We find $\sim$10k comments (0.6\% of
all comments) tagged as \ac{NSFW} that appear \emph{only} in an authenticated crawl's results
with \ac{NSFW}-viewing enabled compared to an unauthenticated baseline crawl.
Although the mechanism for a comment being classified as ``offensive'' is opaque
to us, we similarly discover $\sim$8k (0.5\% of all comments) labeled
``offensive'' by \dis. Posts can be ``reported'' by users, 
which informs the reporting individual that the content will be
``reviewed as soon as possible.'' This mechanism is one hypothesis for
how ``offensive'' comments become labeled; alternatively, \dis may attempt to
automatically label comments based on the presence of certain words in the text.

In order to ensure
that we did not obtain false positive classifications due to HTTP timeouts and other errors
introduced from the crawling framework, we perform two validation steps. First,
we keep track of any \ac{URL} during our crawl that received a timeout error and
mark those \acp{URL} for re-crawling at a later time. We repeat this process
until all pages have been successfully parsed by our framework.  Second, we select a random sample of 100
\ac{NSFW} and ``offensive'' comments, and perform a manual validation to ensure that the comment
only appears when authenticated and with the proper settings enabled. All 100
comments we manually verified were correctly classified as \ac{NSFW} or
``offensive'', although
several were posted both as \ac{NSFW} and also without the label. Because a user
cannot see even their own \ac{NSFW}-labeled comments if they do not have this
visibility setting enabled, it is possible that these duplicate posts occurred when
a user posted a comment with the \ac{NSFW} label enabled but did not see it
appear, and re-posted without the label believing their first post was
unsuccessful.

\begin{figure}[t]
\centering
  \includegraphics[width=\columnwidth]{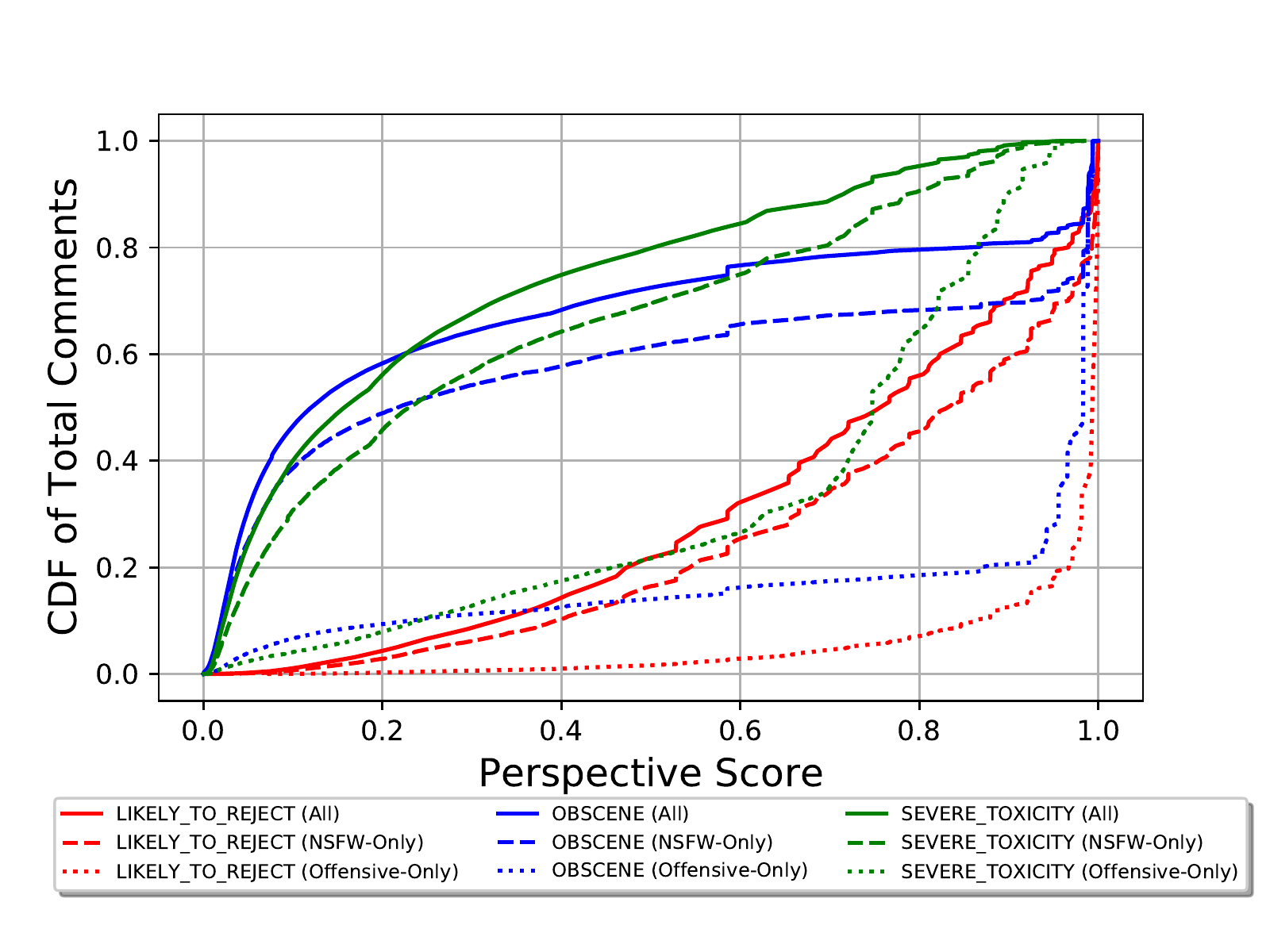}
\vspace{-3mm}
  \caption{\ac{NSFW}, Offensive, and Aggregate Comments Comparison}
 \label{fig:nsfw_toxicity}
\vspace{-3mm}
\end{figure}

We find that \ac{NSFW} content is more toxic than the standard comments and
replies posted to \dis, and ``offensive'' comments are much more so. Figure~\ref{fig:nsfw_toxicity} is a CDF of the 
Perspective scores attributed to \ac{NSFW} and ``offensive'' comments vs the entire comment
population in the Perspective categories of ``\texttt{OBSCENE}'',
``\texttt{SEVERE\_TOXICITY}'', and content likely to be rejected by the New York Times' moderation
section (``\texttt{LIKELY\_TO\_REJECT}''.)  In all three categories, we find the
``offensive''
content to be significantly more extreme than both the unlabeled comments and
replies and the \ac{NSFW} content. For instance, 80\% of the
``offensive'' comments score $>0.95$ in the \texttt{LIKELY\_TO\_REJECT}
category, whereas only 25\% of \ac{NSFW} comments and $< 20\%$ of all comments
score this high. The \ac{NSFW} content is also more extreme than the aggregate
\dis comments, but to a lesser degree than the ``offensive'' comments.
This indicates that users are correctly using the \ac{NSFW} label on
some of the more 
extreme content on \dis, and that the ``offensive'' labeling mechanism captures the most
radical and toxic content. However, because $\sim$85\% of users do not have
the \ac{NSFW} or ``offensive''
view-settings enabled, including both Gab's founder and CTO at the time we ingested 
user settings in \S\ref{sec:user_results}, these comments act as
a network within a social network, where the most extreme content on the
platform appears.

\textbf{Takeaways: } \ac{NSFW} and ``offensive'' content comprises a small fraction of the total
\dis comment corpus; however, their Perspective scores indicate that this content
is substantially more extreme than non-tagged content, and is not
visible to most \dis users with their current settings. The labeling mechanism
for ``offensive'' comments, while unknown, captures the most extreme content on
\dis, which is substantially more toxic than the total \dis comments in aggregate.

\begin{figure}[t] 
  \centering
  \includegraphics[width=0.9\columnwidth]{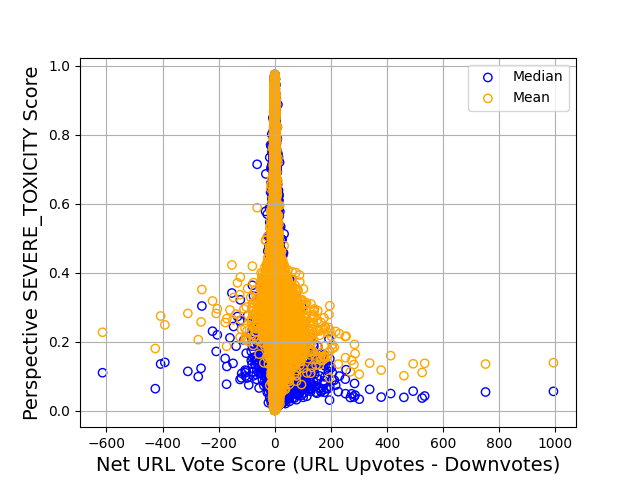} 
  \vspace{-3mm}
  \caption{\texttt{SEVERE\_TOXICITY} Score Compared to \ac{URL} Net \dis Vote Score}
\label{fig:vote-toxicity}
  \vspace{-3mm}
\end{figure}

\subsubsection{Are \acp{URL} with toxic comments up- or down-voted?}
\dis allows authenticated users to cast votes on the \ac{URL} its members have
commented upon by clicking a ``thumbs up'' or ``thumbs down'' button. We collect
this data for 588k \acp{URL} in our crawl, and compare the net vote score
(upvotes minus downvotes) to the Perspective \texttt{SEVERE\_TOXICITY} scores
for comments on these \acp{URL}. 104k \acp{URL} had a positive net vote score,
64k a negative score, and the majority (420k) had a net score of zero. 581k
(99\%) have a net vote score $n \in (-10,10)$.
Figure~\ref{fig:vote-toxicity} plots the mean and median Perspective
\texttt{SEVERE\_TOXICITY} score for each \ac{URL} with its net vote score. 415k
\acp{URL} have no votes in either direction, contributing to the grouping around
the zero net vote score. The zero vote score content exhibits the highest mean
and median \texttt{SEVERE\_TOXICITY} scores, evidenced by the tall peak of
points around $x = 0$. As the net vote score absolute value increases,
however, the \texttt{SEVERE\_TOXICITY} scores corresponding to those \acp{URL}
decrease. \acp{URL} with negative net vote scores in general having higher
\texttt{SEVERE\_TOXICITY} scores than their positive counterparts. 
One possible explanation for this is that \dis users down vote a 
given URL because they disagree with its content.
Disagreeing with said content is also likely a trigger for toxic speech.

\textbf{Takeaways: } Significantly up- or down-voted content appears to generate
lower comment mean and median toxicities, while comments with net vote scores
near zero garner toxicity scores across the spectrum.

\subsection{Relative Toxicity}
\label{sec:results:toxicity}

Next, we consider whether \dis users are more or less toxic than other users and platforms.

\begin{table}[t]
\caption{Overview of baseline toxicity datasets.}
\label{tbl:toxicity_datasets}
\vspace{-3mm}
\small{
\begin{tabular}{lrr}
\hline
Dataset    & \# comments & \# Dissenter users \\ \hline
NY Times   & 4,995,119   & N/A                \\
Daily Mail & 14,287,096  & N/A                \\
Reddit     & 13,051,561 & 35,718
\end{tabular}
}
\vspace{-3mm}
\end{table}

\subsubsection{Baseline Datasets}

In addition to \dis comments, we construct three additional datasets: 1)~NY Times, 2)~Daily Mail, and 3)~Reddit (summarized in Table~\ref{tbl:toxicity_datasets}).
The NY Times and Daily Mail datasets are comments crawled from their respective sites, acquired from~\cite{zannettou2020measuring}.
We chose these two news outlets for a few reasons.
First, they are both relatively popular on \dis; Daily Mail is the 5th most commented on domain by \dis users (see Table~\ref{table:tlds_domains}) and NY Times is the 21st most popular.
Next, they are on different sides of the political spectrum.
Finally, the Perspective API has a set of models that are trained on NY Times
comment moderator decisions, providing additional insight into how the model
describes the specific environment in which it was trained.

The Reddit dataset includes comments by accounts on Reddit that we believe are likely to controlled by a corresponding \dis user.
The construction of the Reddit dataset is a bit more complicated, but allows us to answer some questions related to how \dis users' behavior differs from their behavior on other, moderated social media platforms.
We construct this dataset by querying Reddit for users matching our
known \dis usernames.
This revealed a large number of Reddit users: more than 56k \dis usernames (56\%) correspond to a registered Reddit account.
Of course, different people might choose the same username on different
platforms, so we do not claim that all 56k of these accounts represent the same person on both platforms.
While it is a near certainty that there are false positives in this construction, especially for particularly short usernames or usernames based on common words, previous work~\cite{newell2016UserMigrationOnline} established a lower bound precision of 0.6 for this type of matching and found it sufficient to describe behavioral trends when studying user migration from Reddit.
With these caveats in mind, for each of the 56k identified Reddit accounts, we
query Pushshift~\cite{baumgartner2020pushshift} for all of the comments they made on Reddit.

\begin{figure}[t!]
  \centering
  \includegraphics[width=0.9\columnwidth]{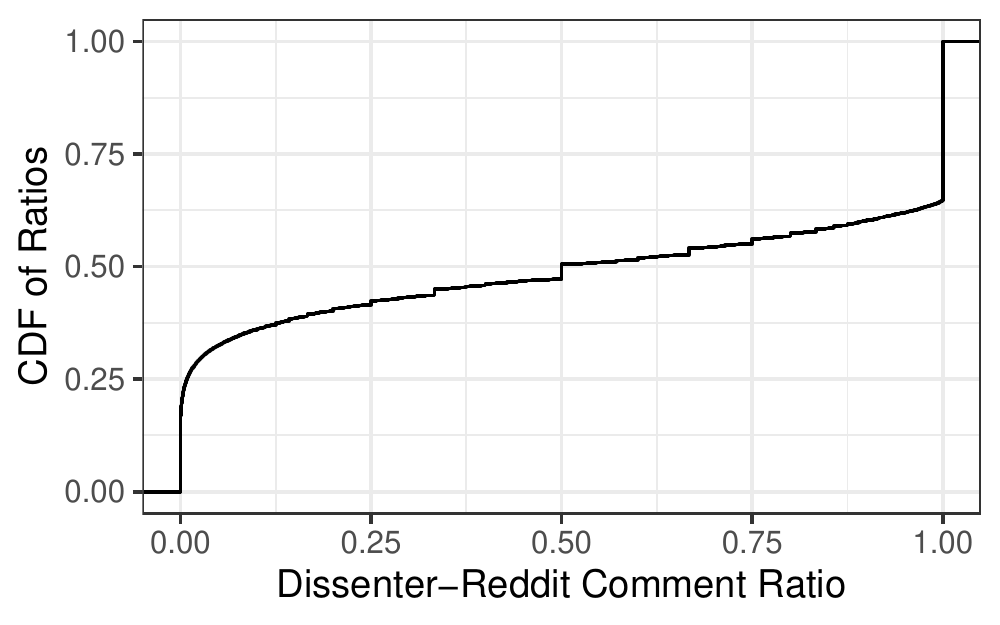}
  \vspace{-3mm}
  \caption{Ratio of Dissenter to Reddit Post Counts}
  \label{fig:reddit_ratio}
  \vspace{-3mm}
\end{figure}

Figure~\ref{fig:reddit_ratio} plots the CDF of user ``comment ratios,'' which is defined as $\frac{d}{d+r}$, where $d$ is the number of posts a user has made on \dis, and $r$ the analogous count on Reddit.
We consider only users that have commented on \emph{at least one} platform so that the ratio is
well-defined; this limits the scope to 31k unique usernames.
There is a roughly even split between which platform has been used more.
The users that have more comments on \dis, however, tend to use that service exclusively, with more than a third having commented \emph{only} on \dis.

\textbf{Takeaways: } A majority of usernames ($\sim$56\%) exist on both \dis and
Reddit. More than a third of users on both platforms post on \dis exclusively, as
opposed to 20\% that post only on Reddit.

\subsubsection{Is the fear of censorship warranted?}

\begin{figure*}[t]
  \begin{subfigure}{.32\textwidth}
    \centering
    \includegraphics[width=.9\columnwidth]{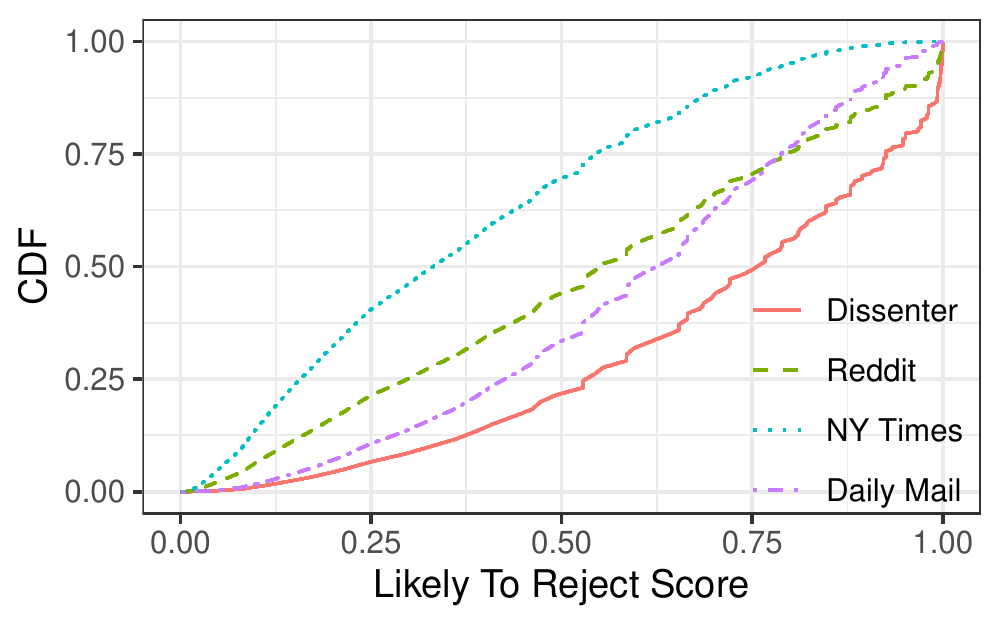} 
    \caption{\texttt{LIKELY\_TO\_REJECT}}
    \label{fig:likely_to_reject}
  \end{subfigure}%
  \begin{subfigure}{.32\textwidth}
    \centering
    \includegraphics[width=.9\columnwidth]{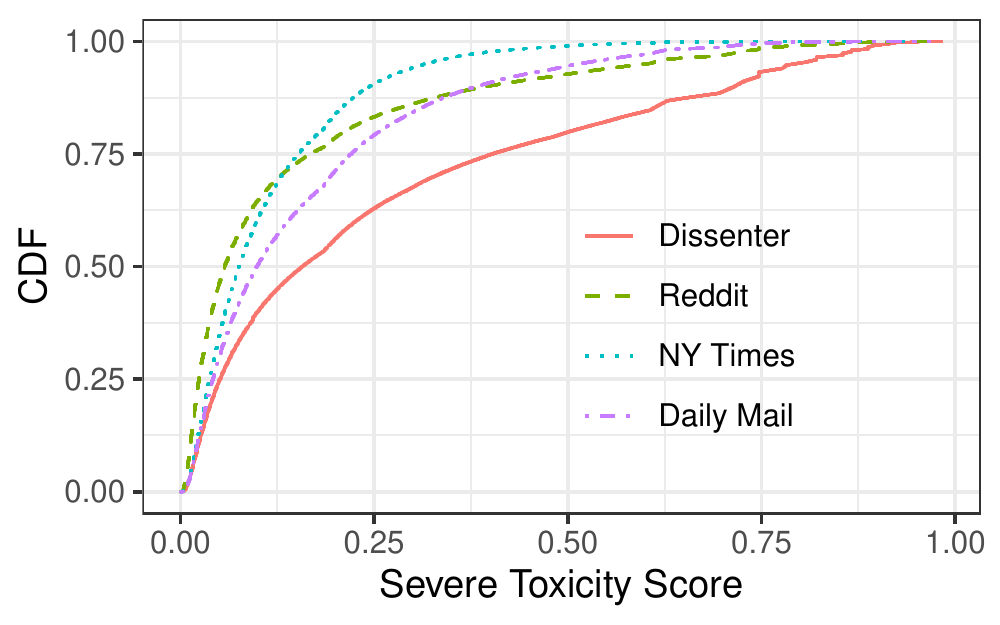} 
    \caption{\texttt{SEVERE\_TOXICITY}}
    \label{fig:severe_toxicity}
  \end{subfigure}%
  \begin{subfigure}{.32\textwidth}
    \centering
    \includegraphics[width=.9\columnwidth]{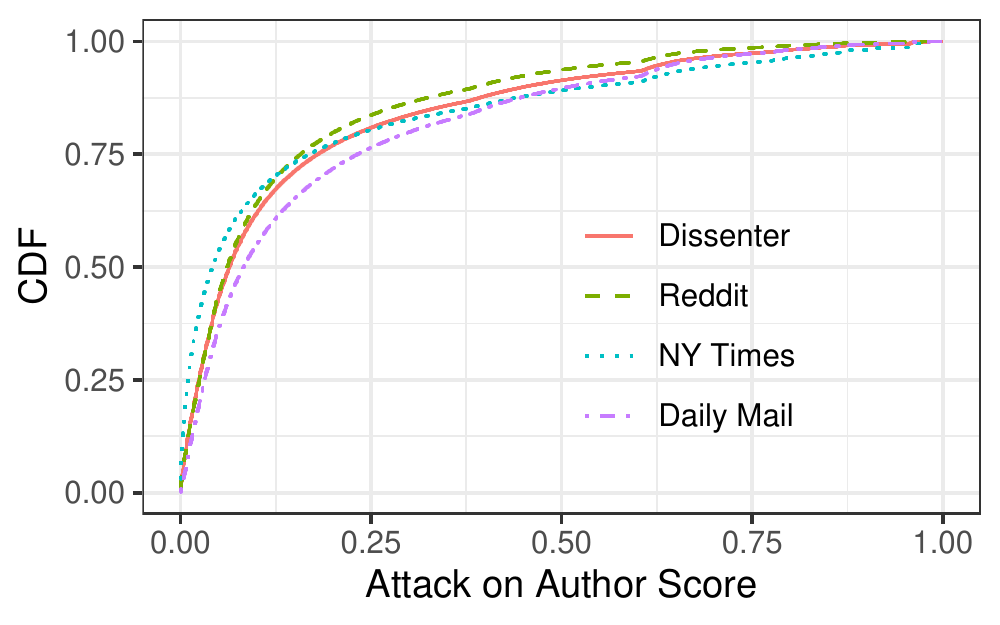} 
    \caption{\texttt{ATTACK\_ON\_AUTHOR}}
    \label{fig:attack-on-author}
  \end{subfigure}%

  \caption{Perspective Model Scores for \dis and Related Datasets}
  \label{fig:perspective_all}
\end{figure*}

One of the motivating factors behind \dis's creation is the belief that moderators are stifling the open discussion of content on their platforms.
The Perspective API provides a model that can help us ascertain whether or not this is the case: the \texttt{LIKELY\_TO\_REJECT} model.
This model is trained on decisions from New York Times comment moderators and provides a score indicating whether or not they would reject a given comment on an article from being published.
Figure~\ref{fig:likely_to_reject} plots the CDF of scores from the \texttt{LIKELY\_TO\_REJECT} model for comments from \dis NY Times, Daily Mail, and \dis users' Reddit accounts.
From the Figure, we see that over 75\% of \dis comments receive a \texttt{LIKELY\_TO\_REJECT} score of 0.50 or more and 50\% of comments receive a score above 0.75.
While the figure makes the different norms on NY Time and Daily Mail quite obvious, \dis comments stand out as being significantly
more likely to reject than comments from other platforms.
We also note that the \dis users' Reddit comments follow a mostly uniform distribution that falls somewhere between the Daily Mail comments and the NY Times comments.
Finally, although we do not show the results for clarity purposes, we found that when looking at just \dis comments on NY Times or Daily Mail URLs the \texttt{LIKELY\_TO\_REJECT} score distributions follow the same shape as the \dis curve in Figure~\ref{fig:likely_to_reject}.

This result indicates that a substantial chunk of \dis comments are indeed considered unsuitable for publishing (at least by NY Times standards) and provides some degree of justification for \dis's motivation.
It further indicates that this behavior might be associated with \dis
itself, since \dis users' Reddit accounts fall somewhere between
comments from NY Times and Daily Mail.
Unfortunately, the \texttt{LIKELY\_TO\_REJECT} model provides no explanation of \emph{why} a comment might have been rejected, but we can get a feeling by looking at the scores from other models.

\textbf{Takeaways: }
The Perspective \texttt{LIKELY\_TO\_REJECT} model indicates
that \dis comments are significantly more likely to be rejected by comment
section moderators,  lending support for \dis's niche in subverting
moderation.

\subsubsection{How toxic are \dis comments?}

The \texttt{SEVERE\_TOXICITY} model offers another window into the type of
comments posted by \dis's user. This Perspective model scores content by its
ability to cause users to feel like they do not want to participate in further
discussion, and is less sensitive to positive uses of profanity (\eg, ``Damn,
that's cool'') than similar toxicity models offered by the Perspective API. A
high \texttt{SEVERE\_TOXICITY} score for a comment indicates a ``very
hateful, aggressive, or disrespectful comment''~\cite{perspective}. 

Like previous work
(\cite{zannettou2018gab,hine2017kek,djuric2015hate}),%
 we compare the amount of toxic hate speech on
\dis to similar platforms. Figure~\ref{fig:severe_toxicity} plots the CDF of
\texttt{SEVERE\_TOXICITY} for our datasets. \dis comments score the
highest in toxicity scores of the four data sources considered; approximately
20\% of \dis comments have a \texttt{SEVERE\_TOXICITY} score $\geq 0.5$,
about double the fraction of Reddit, the nearest dataset. \dis also has
the thickest tail of the four datasets. Roughly 10\% of \dis comments score
0.75 or above, indicating that many comments contain
toxic speech.

\textbf{Takeaways: } \dis comments exhibit substantially higher levels of
toxicity than comments on the other platforms we study. \texttt{SEVERE\_TOXICITY} measures
\emph{very} hateful speech, and is less sensitive to profanity than other
toxicity models.

\subsubsection{Are \dis comments attacking the message or the messenger?}

There is nothing inherently wrong with dissenting opinions.
In fact, dissent is part of healthy debate and discussion.
However, ad hominem attacks serve to stifle productive debate, and also have implications when discussing news in particular.
Figure~\ref{fig:attack-on-author} plots the \texttt{ATTACK\_ON\_AUTHOR} Perspective scores for \dis users, as well as our baseline datasets.
Surprisingly, \dis comments do not a display drastically different tendency to contain an attack on the URL's author.
However, looking at the full distribution here does not reveal the the full picture.

\begin{figure*}[t]
\begin{subfigure}{.5\linewidth}
\centering
    \includegraphics[width=0.8\columnwidth]{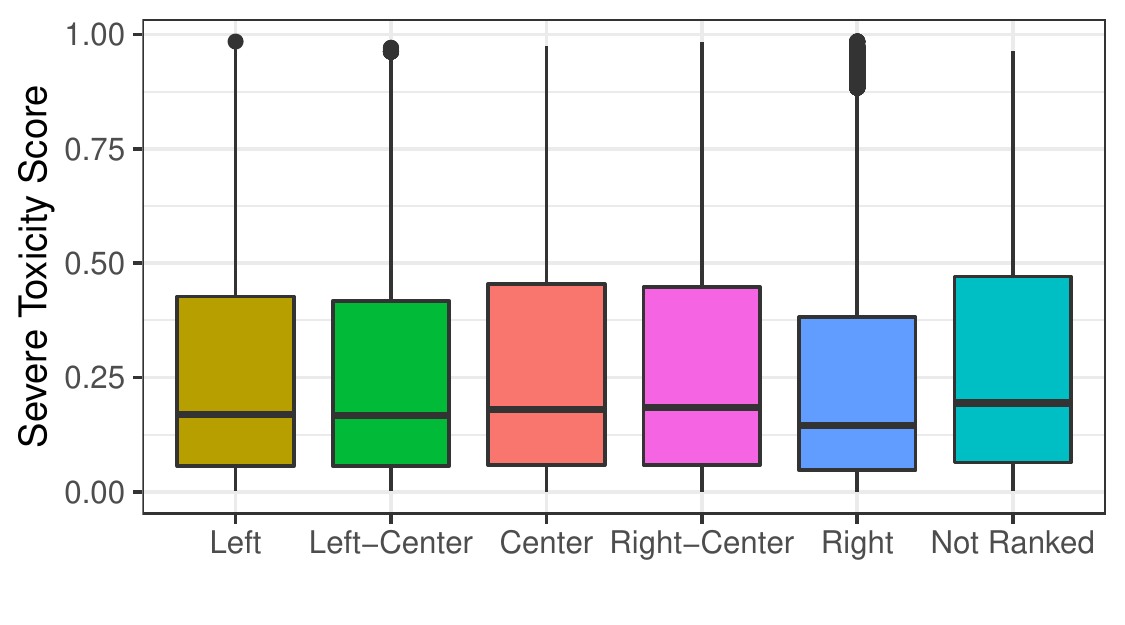} 
    \caption{\texttt{SEVERE\_TOXICITY} Scores}
    \label{fig:allsides_toxicity}
  \end{subfigure}%
  \hfill
\begin{subfigure}{.5\linewidth}
\centering
    \includegraphics[width=0.8\columnwidth]{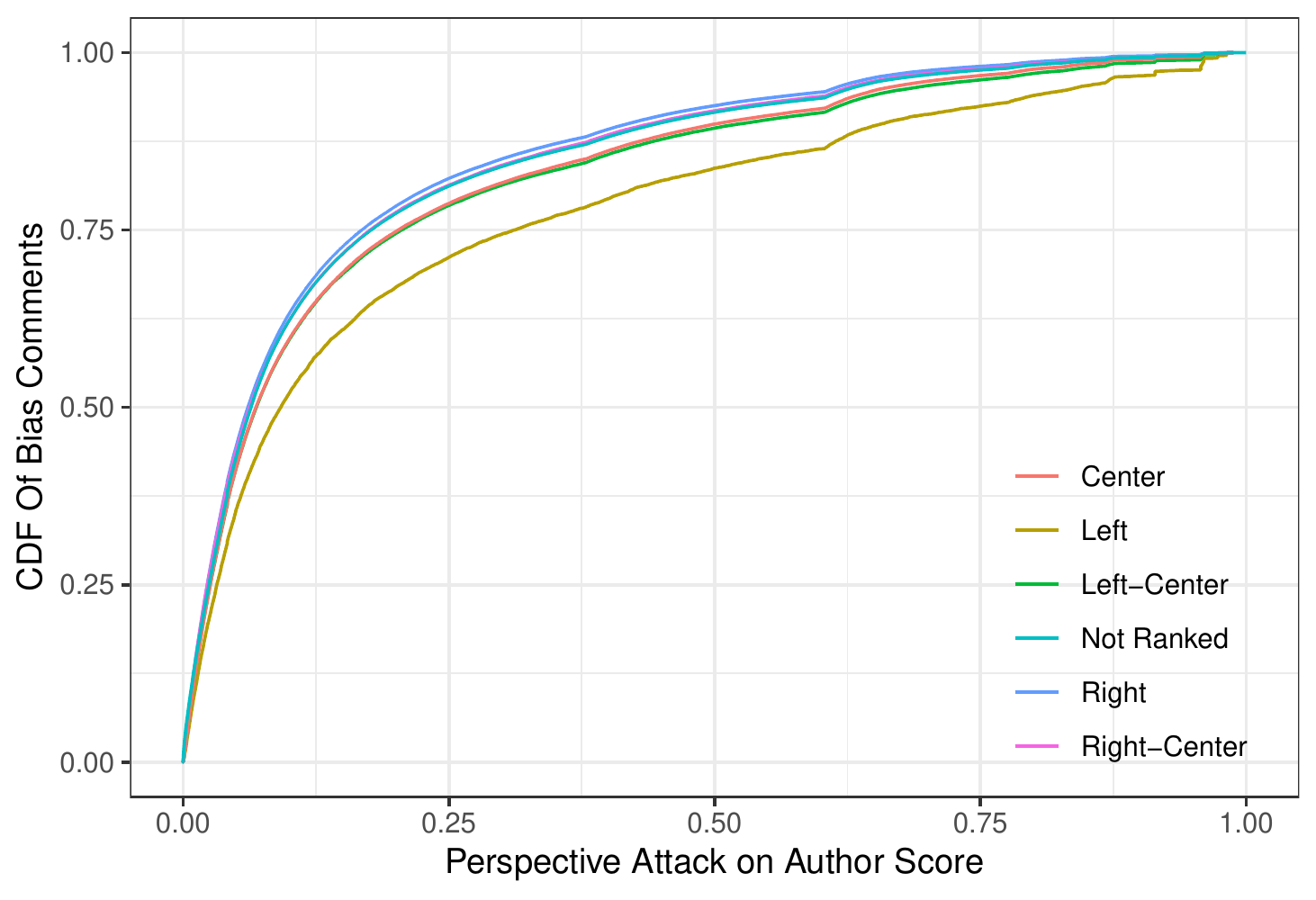}
    \caption{\texttt{ATTACK\_ON\_AUTHOR} Scores}
    \label{fig:allsides_author_attack}
  \end{subfigure}%
  \caption{Perspective Scores By Allsides \ac{URL} Bias Classification}
  \label{fig:allsides}
\end{figure*}

To gain insight into the type of content that elicits toxic \dis user
comments, we use the classification of the commented on \acp{URL} we discover
according to the Allsides media bias rating organization~\cite{allsides}.
Allsides uses multiple methodologies for classifying a media outlet's political
bias, and categorizes popular news media organizations as ``left'',
``center-left'', ``center'', ``center-right'', or ``right''-leaning. By design,
Allsides categorizes the bias of mainstream media organizations and journalists
only; therefore, many \acp{URL} that \dis users comment upon do not have an
Allsides bias. For example, 437k unique comments appear on \yt \acp{URL}; \yt,
as a video sharing service, does not have an Allsides bias ranking (intuitively,
users can post either left- or right-leaning content on the platform.)
Similarly, social media sites do not have an Allsides bias ranking either. Of
1.68M unique comments, approximately 1M fall on \acp{URL} with no Allsides
ranking. The preponderance of these comments 
($\sim$45\%) are on video sharing site \acp{URL}, primarily \yt. Another 110k
are on social media domains, like Twitter, Facebook, and Gab, and 155k
more are on media outlets for which Allsides does not have a bias ranking.

Of the 600k comments on \acp{URL} that have an Allsides bias, we find that
the underlying media bias has a slight, but significant impact (confirmed via two-sample Kolmogorov-Smirnov; all pairs $p < 0.01$) on the toxicity of the comment on that \ac{URL}.
Using the Perspective \texttt{SEVERE\_TOXICITY} scores, we compare the scores according to each
Allsides bias category in Figure~\ref{fig:allsides_toxicity}.
From the Figure, we observe that toxicity tends to be higher for more
center-leaning URLs; right-leaning URLs exhibit lower \texttt{SEVERE\_TOXICITY} than all other bias types.
On the other hand, the \texttt{ATTACK\_ON\_AUTHOR}
scores in Figure~\ref{fig:allsides_author_attack} evince a higher likelihood
that left-leaning content will generate comments that are an attack on the
author of the article than the other Allsides bias rankings (again confirmed significantly different via two-sample KS test with $p < 0.01$).

\textbf{Takeaways: } Comment Perspective scores exhibit a slight, but
statistically significant influence by the underlying media bias.
Interestingly, while \texttt{SEVERE\_TOXICITY} tends to be higher on more center
URLs, \texttt{ATTACK\_ON\_AUTHOR} is higher on left-leaning URLS and decreases
as the media bias moves rightward. 

\subsection{Social Network Analysis}

\subsubsection{Are there clusters of hateful users in the \dis
followers graph?}

We construct the directed \dis social network graph using the data
from~\S\ref{sec:friends}, inclusive of 45,524 \dis users with at
least one comment or reply.  Both the 
in (followers) and out (following) degree distributions 
fit a power law distribution.  
The top three users by number of followers have
10,705, 9,588,
and 8,183 followers, while the three users following the
most other users follow 15,790, 
10,646, and 10,625 others.  Of note, none of 
the top ten highest degree users (in or out) are among the 
most prolific commenters on \dis overall.
This indicates that \dis's user base, while technically a subset of Gab's user base, is not uniformly drawn from Gab's users.
In other words, \dis seems to appeal to a smaller, more niche groups of users.
While it might seem easy to dismiss these users due to their general lack of ``popularity'' on Gab proper, we believe this is dangerous.
Small, extremely niche online communities have repeatedly been shown to harbor hateful and racist activity, have been actively used in disinformation campaigns, and have spawned numerous acts of violence.

\begin{figure*}[t]
\begin{subfigure}{.32\textwidth}
\centering
  \includegraphics[width=\linewidth]{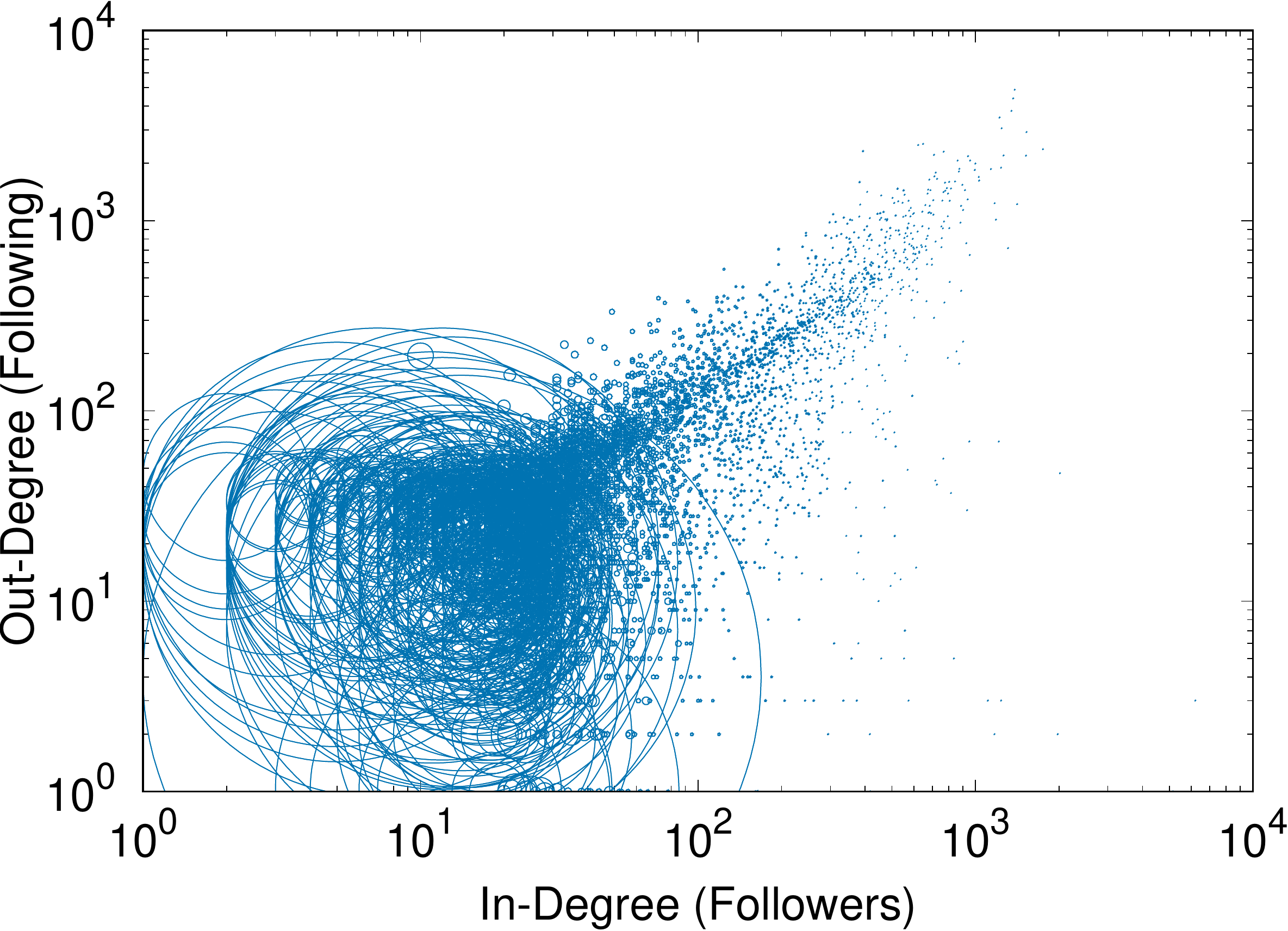}
  \caption{Following vs.\ Followers}
  \label{fig:invsout}
  \end{subfigure}%
\begin{subfigure}{.32\textwidth}
\centering
  \includegraphics[width=\linewidth]{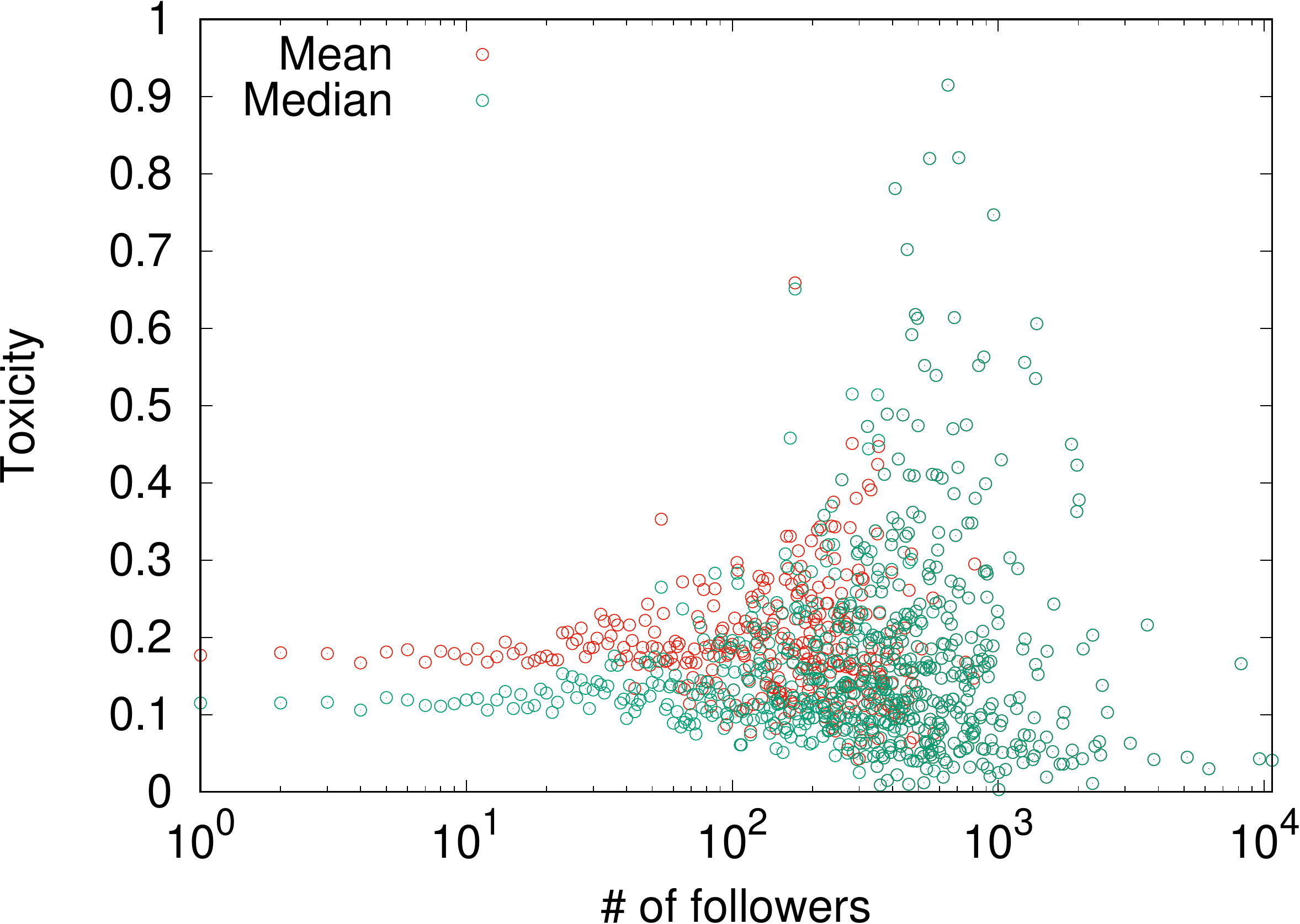}
  \caption{Toxicity vs.\ Followers}
  \label{fig:toxic_v_indegree}
  \end{subfigure}%
\begin{subfigure}{.32\textwidth}
\centering
  \includegraphics[width=\linewidth]{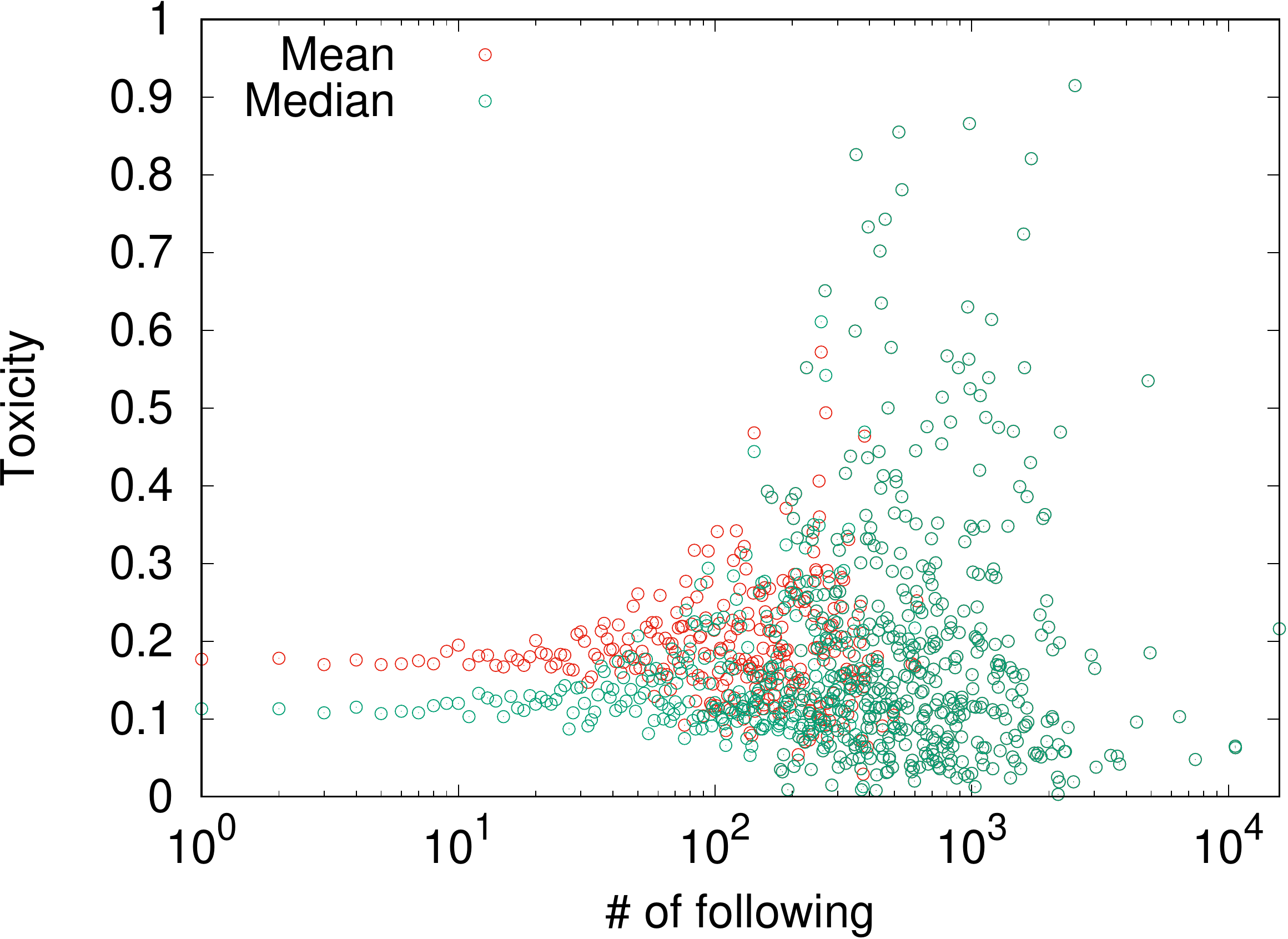}
  \caption{Toxicity vs.\ Following}
  \label{fig:toxic_v_outdegree}
  \end{subfigure}%
\caption{\dis Social Network Analysis}
\label{fig:toxic_v_degree}
\end{figure*}

Figure~\ref{fig:invsout} shows the relationship between number of
users following versus followed.  Fully 15,702 of the users have no
followers and follow no one.  We conjecture that these are Gab users
who tried \dis, but none of their Gab friends or followers are
part of \dis.  In general, however, the number of Dissenters each
user follows is proportional to the number of followers.

Next, we examine the relationship between toxicity and the 
social network graph.  Figure~\ref{fig:toxic_v_indegree} 
shows the mean and median toxicity among \dis users 
for a given number of followers, while
Figure~\ref{fig:toxic_v_outdegree} analyzes the relationship
by number of followed users.  While the toxicity is relatively
low for users that are not well-connected in the graph, 
there are clear outliers with high toxicity and high degree.
Of particular note, the mean is larger than the median for 
small degrees, but is then smaller than the median for higher
degrees -- indicating that the toxicity is skewed depending
on the social network.

Given the macro-level toxicity properties of the social network,
we sought to find the ``hateful core,'' \ie clusters of users
with high toxicity that connect to other users also with high
toxicity.  To find this core, we induce a subgraph on our social
network that includes users $a$ and $b$ iff: i) $a$ and $b$
are mutual followers; ii) $a$ has posted $\ge100$ comments or
replies; iii) $a$'s median comment toxicity is $\ge0.3$.  The
restriction to at least 100 comments is to ensure that the 
user is active; many users have surprisingly high
degree, but with few (or only one) toxic messages.

The resulting hateful core consists of only 42 users, with
six connected components.  There is one large connected component, 
with 32 interconnected users.  While 18 of the usernames have
an active account on Twitter, we find seven of the 32 whose
Twitter accounts were suspended at some point.  The users with
active Twitter accounts largely exhibited self-professed radical
beliefs in their Twitter profiles or posts.  Thus, it appears
that the most connected, hateful users in \dis represent 
both users than have been banned from Twitter, those who use
both platforms, as well as users using \dis and not Twitter.

\textbf{Takeaways: } We examine the induced social network of \dis users on the
Gab social network, and discover a small ``hateful core'' -- a cluster of users
that are active on \dis and routinely post highly toxic content. We find that a
sizable number of these users with Twitter accounts have been suspended. 
\section{Case Studies}

\begin{figure}[t]
 \centering
  \includegraphics[width=0.9\columnwidth]{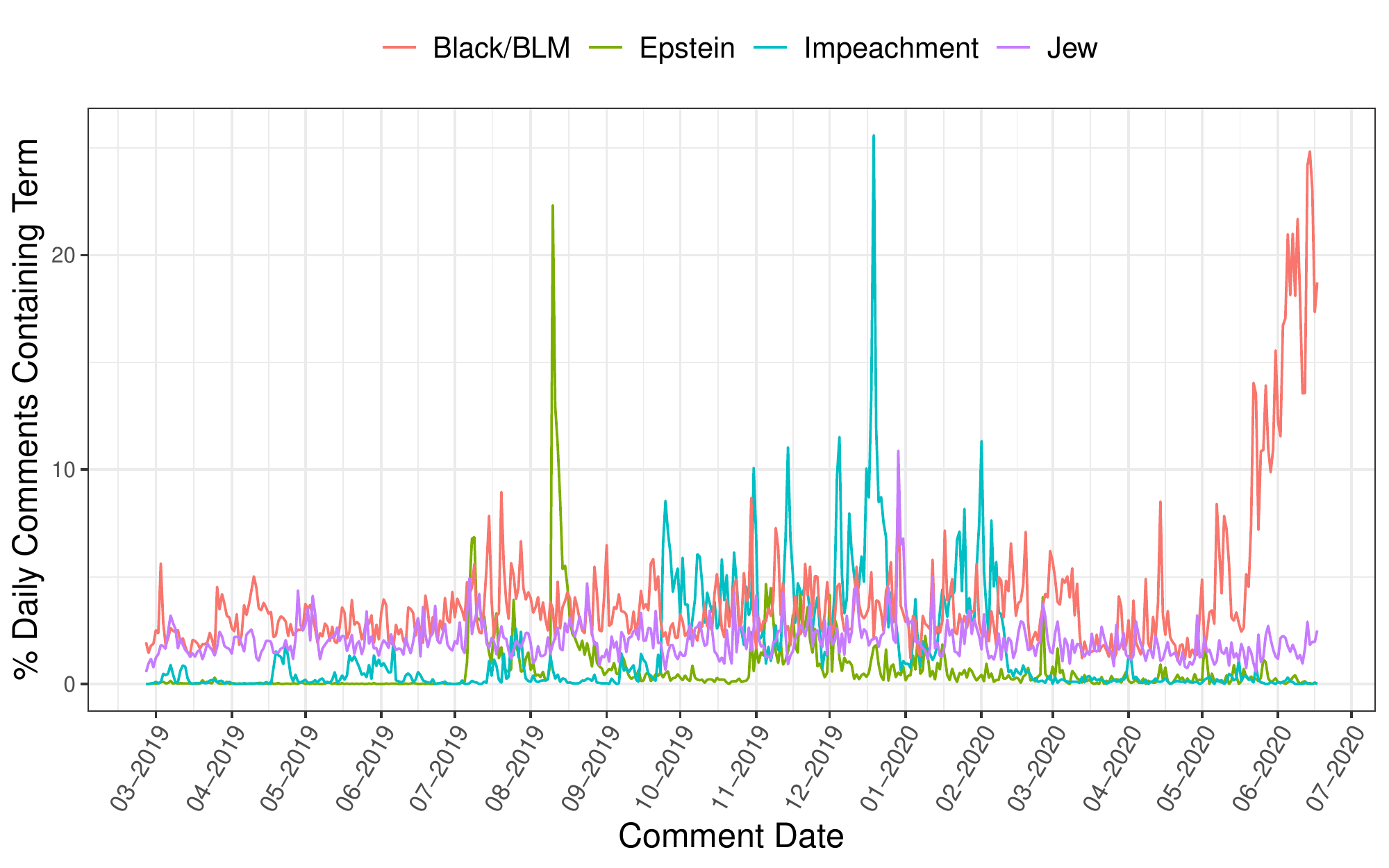}
 \vspace{-3mm}
  \caption{Incidence of Selected Terms in Comments and URLs Over Time}
 \label{fig:keywords}
 \vspace{-3mm}
\end{figure}

The topics \dis users comment about are subject both to the prevailing issues of
the day and to the interests and biases of the user base. This results in
short-term fluctuations in comment themes intermixed with more stable, long-term
topic trends. In this section, we highlight two major events that are
reflected in the \dis data, one term that has elicited steady interest
throughout \dis's existence, and a final topic that exhibits both short- and
long-term community interest behaviors. In all four instances, we consider the
terms' incidence in either a \dis comment or reply, or within the \ac{URL}
string itself. We include the \ac{URL} as part of the analysis in order to
capture comments about content that appears in the \ac{URL} but do not
reference the topic in the comment body. Figure~\ref{fig:keywords} displays the
incidence of each term as a percentage of the total daily comments from
February 2019 to June 2020.  This data
incorporates an additional two months of \dis comments that we crawled in order
to gain insight into the final topic, which occurred subsequent to manuscript
submission.

The first major news event reflected in Figure~\ref{fig:keywords} is the arrest
and subsequent death of Jeffery Epstein, an American financier charged with sex
trafficking minor victims on July 8, 2019~\cite{epsteinArrest}. While his
arrest results in $\sim$7\% of \dis comments referencing
``Epstein'', his death a month later on August 10, 2019 is responsible for a
spike of $\sim$22\% of all \dis comments.%

Although the term ``impeachment'' occurs sporadically throughout \dis's first
six months, it first crosses the threshold of $5\%$ of daily comments in late
September, 2019. This spike coincides with the release of a whistleblower
complaint alleging the US President misused his power to solicit interference in
the 2020 Presidential Election~\cite{whistleblower}. Following this
initial jump,
``impeachment'' occurs at an elevated rate throughout the US
House's impeachment inquiry, reaching a peak of 26\% on December 19, 2019, the
day after the
House voted to impeach the President~\cite{impeached}. 
Another period of increased incidence of ``impeachment'' occurs from mid-January
to early February 2020, coinciding with the impeachment trial and acquittal in
the US Senate.  Subsequent to this event, the term ``impeachment'' occurs in
less than $1\%$ of comments.

In contrast to the terms ``Epstein'' and ``impeachment'', which track closely with
two major US news events, the term ``Jew'' appears with regularity throughout
\dis's existence. Figure~\ref{fig:keywords} shows that ``Jew'' appears in
between 1\% and 5\% of \acp{URL} and comments consistently, breaking above 5\%
once in the final days of 2019, when the term briefly occurred in more than 10\%
of the daily totals. This spike followed a mass stabbing event during Hanukkah in
which five people were wounded at a rabbi's home in New York~\cite{stabbing}.
Despite this singular small peak, the term ``Jew'' represents a long-term
community interest, receiving a sustained non-trivial percentage of daily
comments even in the absence of relevant current events. Previous
work~\cite{zannettou2020quantitative} noted that Gab, \dis's parent social
network, exhibits a high degree of antisemitism. 

Finally, Figure~\ref{fig:keywords} shows the aggregate frequency with which the
words ``Black'', ``BLM'', and a racial slur (not shown) for African-Americans
occurs in daily comments. These terms exhibit the same behavior of ``Jew''
throughout most of \dis's timespan, albeit typically several percentage points
more frequent, and with more spikes throughout the time period. However,
following the death of George Floyd while in police custody in Minneapolis,
Minnesota~\cite{floyd}, which served as a catalyst for widespread protests
throughout the US~\cite{blm} and the world~\cite{worldblm}, the percentage of
comments referencing these terms drastically increased from approximately 5\% of
all comments in March, 2020, to nearly 25\% of daily comments in mid-June, 2020.
This shows that while ``Black'' and associated terms are of long-term community
interest, the frequency of these terms may additionally be propelled by current
events, increasing their appearance in \dis comments.

\section{Related Work}
\label{sec:related}

\textbf{Hate Speech and Toxicity in User Comments. }
Zannettou \etal's study ``What is Gab''~\cite{zannettou2018gab} is most closely
related to this work, as it examines \dis's parent social network. The authors
find that Gab attracts users from the ``alt-right'' and fringe conspiracy
communities and that hate speech is prevalent on the site. As~\cite{zannettou2018gab}
occurred more than a year before \dis's launch, it does not study the \dis
comment aggregation system. Lima \etal~\cite{lima2018inside} also investigate
hate speech on Gab, as well as \acp{URL} that are posted on Gab.  As with
Zannettou's work, \cite{lima2018inside} predates \dis's launch in 2019. Much
prior work exists in detecting hate speech occurrences in online social
networks. Hine \etal study \texttt{/pol/}, a community on the discussion-board
website 4chan in~\cite{hine2017kek}. Among many characterizations of the
discussion board, they attempt to quantify the amount of hate on within the
community, and also identify the \acp{URL} that users post in \texttt{/pol/}
comments. However, unlike \dis, \texttt{/pol/} posts need not include or relate
to a \ac{URL}, and the toxicity of comments vis-\`{a}-vis the underlying
\ac{URL} is not studied. Djuric \etal study the prevalence of hate speech in
Yahoo Finance user comments by using a neural language model trained on
low-dimensional text embeddings of the comments~\cite{djuric2015hate}. 

\textbf{Censorship. }
\dis exists to circumvent user-submitted comment moderation policies, or the
inability to comment at all -- in other words, \emph{censorship} by Big
Tech. Web censorship and censorship detection has a vast body of work; recently,
Yadav \etal \cite{10.1145/3278532.3278555} studied web censorship mechanisms
employed by Indian ISPs. Proxies are often employed to anonymize web traffic and
defeat censorship efforts, and toward understanding the universe of free web
proxies, Perino \etal develop a distributed active and passive measurement system to
measure proxy performance~\cite{perino2018proxytorrent}, which they characterize
in a longitudinal study~\cite{perino2019long}.
In~\cite{10.1145/3355369.3355572}, Raman \etal discuss Mastodon, a decentralized
web microblogging platform that has been forked by Gab in order to avoid being
deplatformed by a single service provider. The pressures toward increasing
centralization that Raman \etal identify could negate the benefit
Gab gains from using a decentralized platform, \eg increased availability,
difficulty to censor, and resilience to outages. Because \dis users are a subset
of Gab's, \dis is also impacted by this trend.
\section{Conclusions and Future Work}
\label{sec:conclusions}

\dis is an approach to evading content provider censorship by
decoupling the comment system from the underlying content; as opposed to network-level
tools (\eg Tor) that address censorship and privacy at the protocol level, \dis
is concerned entirely with the application layer.
Unlike more general social media (\eg Twitter, Reddit, and even Gab) 
user activity on \dis is clearly bound to off-site content. 

Like its parent social network, Gab, \dis claims to support its users' right to
free speech; in practice, this manifests itself in 
toxic content.
In fact, our study
shows that \dis contains more hate-speech than
prior work on Gab~\cite{zannettou2018gab}. However repulsive the content on
\dis, it clearly fills a need for its user base: \dis comments score higher on a
machine learning model trained to classify comments as ``likely to be
rejected by a content moderator'' than comments from any other data source we
studied.

Several additional interesting security and privacy properties of \dis
bear discussion.  First, a small number of \dis comments on non-HTTP(S)
scheme URLs are the result of a user using \dis to view content on
their local file system.  These comments leak information about the
user's file system and the content they have downloaded.  More curious
are \dis comments on web browser start pages and tabs, \eg
``\texttt{chrome://startpage/}''.  
Indeed, any URL is a potential anchor for a \dis comment thread,
suggesting the possibility for a potential form of covert channel, a
hidden conversation within a hidden conversation.  The URL need not
exist, can use any arbitrary scheme, and could be shared among users
wishing to engage in a hidden conversation within the \dis platform.
As we cannot easily differentiate between web URLs that are
no longer responsive versus intentionally fictitious URLs, we leave this
investigation for future research.

Second, a concerning aspect of \dis is the inability of a content
owner to prevent discussion on their content within the \dis
framework.
It is not readily possible to block the \dis browser
via traditional fingerprinting techniques as it is built on the Brave
codebase and does not report a distinct user-agent string.
Interestingly, a proactive defense may discourage or even break the
current \dis model.  A content producer could preemptively post
comments within \dis for the content they own to overwhelm the
conversation with positive comments.  
This has the tangible effect of content publishers being able to potentially affect the way that Dissenter discussions might go.
Such proactive approaches are
important to investigate further in an environment of active
de-platforming.

\ifx\jeremy
\jbnote{begin righteous rant}
The measurements community should be particularly interested in what \dis represents moving forward.
It is without question that computer scientists not only have a role to play in addressing online safety concerns, but also a \emph{responsibility}.
In some ways, \dis is a relative success for much of the measurements and networking community.
Much of our collective effort has gone into understanding how to build decentralized protocols like Mastodon~\cite{10.1145/3355369.3355572} and learning about methods of tracking and evasion from online advertisers~\cite{whatever,wahtever,whatever,whatever}.
Our commitment towards privacy, robustness of the Web, and open source engineering laid the ground work for production quality applications that integrate cutting edge techniques to improve people's online experience.
But at the same time, our efforts have been co-opted.
\dis essentially circumvented de-platforming attempts by co-opting some of those efforts.

We further argue that future research in the measurements community is uniquely positioned to have meaningful impact in addressing societal problems toxicity and hate speech, and this is exemplified by our study of \dis.
While we do not claim to be policy makers, lawyers, sociologists, psychologists, communications experts, etc., we \emph{do} claim that our expertise lies in the general space of data-driven understanding of large-scale, empirical problems.
As the world faces an unprecedented pandemic, as we have witness vastly increased levels of violence directly tied to our contributions to the information age, as we watch questions of online censorship and de-platforming becoming world-wide political issues, we should not sit idle.
Instead, we should leverage our abilities and focus on providing policy makers et al. with a comprehensive set of methodologies, experimentation, and results of what is \emph{really} going on online.

\jbnote{end righteous rant}
\else
Finally, the community should be particularly interested in what \dis represents moving forward.
It is without question that computer scientists not only have a role to play in addressing online safety concerns, but also a \emph{responsibility}.
Further, future research in the measurements community is uniquely
positioned to have meaningful impact in addressing societal problems
of toxicity and hate speech.
The work of policy makers and other experts depends on
a data-driven understanding of these emergent platforms and networks.
As the world faces unprecedented challenges, including
increased levels of violence directly tied to the modern information age
and fierce debate over censorship and de-platforming,
it is incumbent on us to provide methodologies, experimentation, and
results to
understand what is \emph{really} going on online.
\fi

\section{Ethical Considerations}
\label{appendix:ethics}

Considering \dis exists as a service designed to circumvent the moderation
policies and discontinued comment sections of content providers, a further
discussion of the ethical implications of our study is warranted. 

By its nature, \dis users employ pseudonyms and use the platform to
converse anonymously. However, all of the information that we collected in this
study is publicly available and is in fact meant for consumption by normal
\dis users. While we acknowledge that public availability is not a panacea for 
the potential misuse of data, we believe that the unmitigated risk to society
that racist and hateful communities like \dis can pose outweigh the potential
harm that could come from our study. Specifically, we believe that the type of
understanding gained from the present study is not only sufficient but 
\emph{necessary} to designing solutions that guard against the exploitation of
the Internet to harm others.

Some users may purposefully or
inadvertently post personally identifiable information (PII).  While
our analyses do not depend on any PII, we requested a determination
from our Institutional Review Board (IRB) to ensure we were acting
ethically.  Our IRB found that the data we analyze is from publicly
available Internet posts where there is no reasonable expectation of
privacy, and that our research methods support beneficence and respect
for persons.  Finally, none of our work violates \dis's terms of  
use policies.

\section*{Acknowledgements}

We thank Ben Zhao for shepherding and the anonymous reviewers for their
insightful feedback. This work was partially supported by a Content Policy
Research on Social Media Platforms award from Facebook Research. Views and
conclusions are those of the authors and should not be interpreted as
representing the official policies or position of the U.S. government.

\clearpage
\newpage
\small{
\bibliographystyle{ACM-Reference-Format}
\bibliography{dissent}
}

\begin{acronym}
  \acro{CTO}{Chief Technology Officer}
  \acro{URL}{Uniform Resource Locator}
  \acro{NSFW}{Not Safe For Work}
  \acro{TLD}{Top-Level Domain}
  \acro{ccTLD}{Country-Code \ac{TLD}}
\end{acronym}

\end{document}